\documentclass[pdflatex,sn-basic]{sn-jnl}% Basic Springer Nature Reference Style/Chemistry Reference Style

\usepackage{graphicx}%
\usepackage{multirow}%
\usepackage{amsmath,amssymb,amsfonts}%
\usepackage{amsthm}%
\usepackage{mathrsfs}%
\usepackage[title]{appendix}%
\usepackage{xcolor}%
\usepackage{textcomp}%
\usepackage{manyfoot}%
\usepackage{booktabs}%
\usepackage{algorithm}%
\usepackage{algorithmicx}%
\usepackage{algpseudocode}%
\usepackage{listings}%

\usepackage{bm}

\DeclareMathOperator*{\argmax}{arg\,max} % thin space, limits underneath in displays
\newcommand{\fil}[1]{\widehat{#1}}
\newcommand{\dt}[1]{\frac{\partial#1}{\partial t}}
\newcommand{\dxi}[1]{\frac{\partial#1}{\partial x_i}}
\newcommand{\dxj}[1]{\frac{\partial#1}{\partial x_j}}
\newcommand{\dxjj}[1]{\frac{\partial^2#1}{\partial x_j \partial x_j}}
\theoremstyle{thmstyleone}%
\theoremstyle{thmstyletwo}%

\theoremstyle{thmstylethree}%

\begin{document}

\title[MFBO Wake Steering]{Multi-Fidelity Bayesian Optimisation of Wind Farm Wake Steering using Wake Models and Large Eddy Simulations}

\author*[1]{\fnm{Andrew} \sur{Mole}}\email{a.mole@imperial.ac.uk}
\author[2]{\fnm{Sylvain} \sur{Laizet}}\email{s.laizet@imperial.ac.uk}

\affil*[1]{\orgdiv{Department of Aeronautics}, \orgname{Imperial College London}, \orgaddress{\street{Street}, \city{London}, \postcode{SW7 2AZ}, \state{State}, \country{UK}}}


\abstract{Improving the power output from wind farms is vital in transitioning to renewable electricity generation.
However, in wind farms, wind turbines often operate in the wake of other turbines, leading to a reduction in the wind speed and the resulting power output whilst also increasing fatigue.
By using wake steering strategies to control the wake behind each turbine, the total wind farm power output can be increased.
To find optimal yaw configurations, typically analytical wake models have been utilised to model the interactions between the wind turbines through the flow field.
In this work we show that, for full wind farms, higher-fidelity computational fluid dynamics simulations, in the form of large eddy simulations, are able to find more optimal yaw configurations than analytical wake models.
This is because they capture and exploit more of the physics involved in the interactions between the multiple turbine wakes and the atmospheric boundary layer.
As large eddy simulations are much more expensive to run than analytical wake models, a multi-fidelity Bayesian optimisation framework is introduced.
This implements a multi-fidelity surrogate model, that is able to capture the non-linear relationship between the analytical wake models and the large eddy simulations, and a multi-fidelity acquisition function to determine the configuartion and fidelity of each optimisation iteration.
This allows for fewer configurations to be evaluated with the more expensive large eddy simulations than a single-fidelity optimisation, whilst producing comparable optimisation results.
The same total wind farm power improvements can then be found for a reduced computational cost.}




\maketitle

\section{Introduction}\label{intro}

For the realisation of transitioning to producing electricity through 100\% renewable means, a diverse portfolio of power generation methods are required.
In recent years, wind power has emerged as an important contributor to achieving this goal.
The IEA net-zero road map~\citep{NetZeroRoadmap2023} details a need to triple the capacity of renewable electricity generation by 2030 with wind capacity rising from 75GW in 2022 to 320GW in 2030.
Constraints on the placement of wind turbines, including connecting to the grid and ease of maintenance, result in many wind turbines being clustered together in wind farms.
These factors, combined with the need to increase the total capacity, mean that new wind farms consist of a greater number of larger wind turbines.
In these scenarios, the energy output of a wind turbine is coupled to those around it through the dynamics of the air flow.
This contributes to the reduced efficiency of the wind turbines through the wake effect.
With wind turbines located in the wake of other wind turbines, they experience lower velocity and more turbulent wind conditions resulting in a reduced power output and higher wear.
With upcoming wind farms consisting of a larger number of wind turbines, more of the turbines will be positioned in the wake of other turbines.

The effect of this can be reduced through the optimising the placement of the wind turbines within the wind farm~\citep{mosettiOptimizationWindTurbine1994,wanWindFarmMicrositing2012,serranogonzalezReviewRecentDevelopments2014,al-addousSignificanceWindTurbines2020}.
Often referred to as micro-sitting, this method finds an optimum wind farm layout considering the probability distribution of wind conditions that will be experienced by the farm.
However, the range of wind conditions that a wind farm experiences makes finding a globally optimum wind farm layout challenging.
During the operation of the wind farm, the effect of wake losses can be reduced through a globalised control of the wind turbines in the wind farm.
The controllable wind turbine parameters such as the yaw, pitch and induction factors can be used to control the strength and direction of the wind turbine wakes~\citep{flemingSimulationComparisonWake2015}.
Through controlling the individual turbines wakes, the wake losses experienced by other turbines can be reduced.
Although the control may reduce the power output of the individual turbine, the goal is to increase the total power output of the wind farm~\citep{simleyResultsWakesteeringExperiment2021}.
Finding the optimal control parameters leads to an optimisation problem for maximising the farm power output.
Identifying advanced wind farm control systems was highlighted as an area requiring attention in~\citet{veersGrandChallengesWind2022a} and~\citet{veersGrandChallengesDesign2023}.

As the controls and power output of the wind turbines are coupled to each other through the flow field, the most direct way to evaluate this is through solving the Navier-Stokes equations with direct numerical simulations (DNS).
However, the high Reynolds numbers of the flow and consequentially the large range of scales present makes the use of DNS computationally unattainable.
Therefore, a degree of modelling must be introduced to make the problem tractable.
A variety of different fidelities of methods exist for modelling the wakes in a wind farm.
These range from analytical wake models through Reynolds averaged Navier-Stokes (RANS) calculations and to large eddy simulations (LES).
Analytical wake models consist of an analytical approximation of the velocity deficit created by a wind turbine and superimposes these onto the mean flow field~\citep{jimenezApplicationTechniqueCharacterize2010a,bastankhahExperimentalTheoreticalStudy2016a,kingControlorientedModelSecondary2021}.
Optimisation problems, such as micro-sitting or wake-steering, have typically been addressed using analytical wake models due to their computational efficiency allowing for many different configurations to be tested when searching for an optimum.
However, as the wake models are missing important physical flow features, they may find an optimum that does not reflect the true optimum of the real wind farm.
The computational cost of using high-fidelity methods in optimisations means that the number of studies that have investigated this is more limited~\citep{kingOptimizationWindPlant2017,bempedelisDatadrivenOptimisationWind,asmuthHowFastFast2023}.
By capturing the complexities of non-linear and unsteady fluid dynamics, LES delivers solutions closer to the true optimum.

A limited number of studies have investigated combining low- and high-fidelities for wind farm investigations.
The work of~\citet{bempedelisDatadrivenOptimisationWind} implements an extreme informed LES BO where the optimisation results from low-fidelity analytical wake models are used to provide the initial sampling locations of a high-fidelity BO using LES\@.
Although this method provides an informed starting point for the high-fidelity BO, the lack of modelling between the fidelities means that it may hamper the convergence of the high-fidelity BO, trapping in local optimum rather than finding the global optimum.
The work of~\citet{anagnostopoulosAcceleratedWindFarm2023} combines low- and high-fidelity wake models through transfer learning of neural networks to generate a new surrogate model.
Although this method could be extended to include data from truly high-fidelity simulations, the methodology requires a large training data set before optimisation.
Low-fidelity wake models and high-fidelity LES data is combined in~\citet{hoekPredictingBenefitWake2020} using Gaussian process (GP) regression.
However, both fidelities of data are combined into one dataset before being used to create the model with no modifications made for handling the differing fidelities.
The work of~\citet{anderssonRealtimeOptimizationWind2020a,anderssonAdaptationEngineeringWake2020,anderssonGaussianProcessesModifier2020} utilises GPs in a modifier adaption framework to model the mismatch between an analytical wake model and the actual plant measurements.
Although the method models the relationship between the fidelities and obtains good results for few turbines, the requirement to train the GP before performing the optimisation, and the reliance on modelling directly an estimation of the error, means that the performance drops off sharply when analysing wind farms with more than four turbines.

A multi-fidelity Bayesian optimisation (BO) approach is used in~\citet{quickMultifidelityMultiobjectiveOptimization2022} to find wake steering strategies.
This approach removes the need to pre-train the GP model with sufficient data before running the optimisation.
However, the multi-fidelity model employed still relies on an additive relationship between the models which limits the complexity of the relationship between the fidelities that can be discovered.
To circumvent this restriction, the fidelities of the models are chosen to be as correlated as possible, with LES used for both fidelities.
This means that there is still a considerable cost associated with the low-fidelity experiments.
The limitation of a linear correlation between the fidelities was addressed in~\citet{kirbyDataDrivenModelling2023} where a non-linear auto-regressive Gaussian process (NARGP) is used to capture the relationship between the fidelities.
They showed that combining wake model and LES data provided an improvement over a single-fidelity approach for large wind farm layouts.
However, the model was only used as a surrogate model and was not extended for use in an optimisation framework.

In this work, we propose a non-linear auto-regressive based multi-fidelity framework based on LES and analytical wake models for optimising the power output of a wind farm using yaw control strategies.
The paper contains a description of the multiple fidelities of wind farm environments (wake models and LES) utilised in this work in Sect.~\ref{sec:environments}.
This is followed by a description of a typical single-fidelity optimisation framework and its extension to multi-fidelity optimisation in Sect.~\ref{sec:bo}.
Section~\ref{sec:results} contains results of the yaw control optimisations, comparing single- and multi-fidelity strategies in both a relatively simple two-turbine layout and in a larger, more realistic wind farm layout.
Finally, Sect.~\ref{sec:conclusions} concludes the paper and presents a summary of the main contributions.

\section{Wind Farm Environments}\label{sec:environments}

In this work, methods of different fidelities are used to evaluate the wind farm power output for given wind farm configurations.
The bulk of the work uses analytical wake models as a low-fidelity wind farm environment and LES as a high-fidelity wind farm environment.
When conducting a BO, these methods are used as a black box, providing a total wind farm power output when supplied with the turbine yaw configurations.

\subsection{Analytical Wake Models}\label{subsec:wake}

In this work, the analytical wake models are implemented using the FLORIS~\citep{floris} (v3.4.1) package, which is a control-focused wind farm simulation software incorporating steady-state engineering wake models.
When provided with initial atmospheric conditions and turbine parameters as inputs, FLORIS models the steady stream-wise velocity deficit created by each turbine in an array.
These stream-wise velocity deficits are combined using a superposition model to generate a final flow field.
Superposition models estimate the combined effects of multiple turbine wakes by combining their individual impacts.
They are needed to efficiently handle the interactions between wakes in large wind farms without having to solve for every possible interaction in detail.
In this study, the sum of the squares freestream superposition model of~\citet{Katic1987} is used to combine the wakes.
This defines the velocity field as
\begin{equation}\label{eq:wake-superposition}
    u = U_0 - \sqrt{\sum_n^N \left(U_0-u_w^n\right)^2}
    \;,
\end{equation}
where $U_0$ is the freestream velocity and $u_w^n$ is the wake velocity induced by the $n$th wind turbine in stand-alone conditions defined by a wake model.
Once a flow field has been generated, look-up tables can be used to convert the wind speed at each turbine's location into power and thrust coefficients.
A range of wake models are available that capture varying features of the turbine wakes.
In this work, the Jensen wake model, the Gaussian wake model, and the Gauss Curl hybrid (GCH) wake model are used, representing a range of fidelities of wake models.

\subsubsection{Jensen Wake Model}
The Jensen model~\citep{jensen1983}, also known as the Park model, calculates the wake velocity deficit as constant across the wake's span-wise cross-section and abruptly returns to freestream conditions outside of the wake.
The wake expands linearly down-stream according to the wake decay constant $k$, which defines the wake growth rate.
In yawed conditions, the wake centerline is deflected according to the Jimenez model~\citep{jimenezApplicationTechniqueCharacterize2010a}, which accounts for the deflection of the wake due to yawed turbine operation.
The wake velocity deficit $\delta u = 1 - u_w/U_0$ at a distance $x$ downstream of the turbine is given by,
\begin{equation}\label{eq:jensen-wake}
    \delta u = \frac{2 a U_0}{(1 + 2 k x / D)^2}
    \;,
\end{equation}
where $a$ is the axial induction factor, $U_0$ is the freestream velocity, $k$ is the wake decay constant, $x$ is the downstream distance, and $D$ is the rotor diameter.
The discontinuous nature of the velocity field produced by this model makes it ill-suited to wake steering applications~\citep{goriSensitivityAnalysisWake2023c}.

\subsubsection{Gaussian Wake Model}
The Gaussian wake model, developed by~\citet{bastankhahNewAnalyticalModel2014}, uses a Gaussian distribution to model the velocity deficit in the span-wise direction.
This ensures a gradual transition from the wake centre to the freestream conditions, providing a more realistic representation of the wake profile.
The velocity deficit $\delta u$ at a point $(x, r)$ downstream is given by,
\begin{equation}\label{eq:gauss-wake}
    \delta u = U_0 \left(1 - \left(1 - \frac{2 a}{(1 + k \frac{x}{D})^2}\right) \exp\left(-\frac{r^2}{2\sigma^2}\right)\right)
    \;,
\end{equation}
where $r$ is the radial distance from the wake centerline, $\sigma$ is the standard deviation of the Gaussian profile, and $k$ is the wake expansion coefficient.
This model provides a continuous and smooth velocity field, making it more suitable for representing wake interactions and control strategies.

\subsubsection{Gauss-Curl Hybrid (GCH) Wake Model}
The GCH model~\citep{kingControlorientedModelSecondary2021} extends the Gaussian model by incorporating additional physical effects such as wake rotation and counter-rotating vortex pairs.
These effects help to accurately model the wakes in yawed conditions and wake steering scenarios.
The GCH model represents the wake deficit using a Gaussian distribution while introducing curl terms to account for the rotational effects.
The velocity deficit $\delta u$ is calculated using the same equation~\ref{eq:gauss-wake} as for the Gaussian model, and the additional rotational velocity components $(u_\theta)$ in the wake are given by,
\begin{equation}\label{eq:gch-wake}
    u_\theta = \Gamma \frac{r}{2 \pi \sigma^2} \exp\left(-\frac{r^2}{2\sigma^2}\right)
    \;,
\end{equation}
where $\Gamma$ is the circulation strength of the wake's rotational component.
This model captures wake asymmetry, yaw-based wake recovery, and secondary steering effects,~\citep{kingControlorientedModelSecondary2021} providing a more accurate representation of the wake under various conditions~\citep{goriSensitivityAnalysisWake2023c}.

In summary, the Jensen, Gaussian, and GCH wake models implemented in FLORIS offer a range of approaches for modeling wind turbine wakes, increasing in fidelity, for wake steering applications.

\subsection{Large Eddy Simulation (LES)}\label{subsec:les}
In order to obtain a high-fidelity prediction of the flow and corresponding power output around a particular wind farm configuration, LES is used.
This work uses the high-fidelity wind farm simulator Winc3d~\citep{deskosWInc3DNovelFramework2020} based on XCompact3D~\citep{bartholomew2020xcompact3d} to solve the incompressible, explicitly filtered Navier Stokes equations.
The obtained governing equations for the conservation of momentum and mass are given by,
\begin{equation}\label{eq:les-ns-2}
\dt{\fil{u_i}} + \frac{1}{2} \left( \fil{u_j} \dxi{u_i} + \dxj{\fil{u_i} \fil{u_j}} \right) = - \frac{1}{\rho} \dxi{\fil{p}} - \dxj{\tau_{ij}} + \nu \dxjj{\fil{u_i}} + \frac{F_i}{\rho}
\;,\quad
(i = 1, 2, 3)
\;,
\end{equation}

\begin{equation}\label{eq:les-ns-1}
\dxj{\fil{u_j}} = 0
\;.
\end{equation}
Here, spatial filtering is denoted by $\fil{\cdot}$ and is applied to the pressure, $p$, and velocity components, $u_i$, with summation implied over the $j$ index.
The fluid properties of the density and kinetic viscosity are denoted by $\rho$ and $\nu$ respectively.
$F_i$ denotes all forces applied to the fluid and $\tau_{ij}$ are the subfilter stresses where $\tau_{ij} = \fil{u_i u_j} - \fil{u_i}\fil{u_j}$.
These stresses are modelled using the standard Smagorinsky model,
\begin{equation}\label{eq:smagorinsky}
\tau_{ij} = -2 (C_s \Delta)^2 \fil{S_{ij}} |\fil{S}|
\;,
\end{equation}
where $\fil{S_{ij}}$ and $|\fil{S}|$ are the filtered strain-rate tensor and its magnitude,
respectively, $\Delta$ is the filter width and $C_s$ is the Smagorinsky coefficient.
In this work, the Smagorinsky coefficient is damped close to the wall using the wall damping function of~\citet{masonStochasticBackscatterLargeeddy1992}.
This defines the Smagorinsky coefficient at a height above the wall $y$ as,
\begin{equation}\label{eq:MTdamp}
C_s = \left( C_0^n + \left[ \kappa \frac{y+y_0}{\Delta} \right]^{-n} \right)^{-1/n}
\;,
\end{equation}
where, the Smagorinsky constant far from the wall $C_0 = 1.4$, the von K\'{a}rm\'{a}n constant, $\kappa=0.4$, the growth parameter $n = 3$, and $y_0$ is the height of the surface roughness.

Winc3d is based on a Cartesian mesh and employs compact finite-difference schemes for differentiation, filtering, and interpolation.
The spatial discretisation is achieved using sixth-order implicit finite-difference schemes.
For the time advancement, an explicit third-order Adams-Bashforth scheme is applied.
The high-order compact finite difference schemes used ensure ``spectral-like'' accuracy while accommodating non-periodic boundary conditions.
The quasi-spectral accuracy of these high-order schemes significantly improves the efficient resolution of small-scale turbulence at a given resolution, offering an advantageous balance between computational cost and accuracy~\citep{Laizet2009}.

To prevent numerical oscillations, a spatial filter is applied in all directions to the velocity field and pressure field at each time step, utilizing a discrete filter based on the tridiagonal filtering method by~\cite{Motheau2016}.
The discrete low-pass filter coefficients are set to achieve the required sixth-order accuracy, and a user-defined coefficient $\alpha_{filter} = 0.49$ is used to filter near the grid scale.
The incompressibility of the velocity field is maintained by solving the pressure Poisson equation in spectral space using three-dimensional fast Fourier transforms (FFTs) and the concept of the modified wave number~\citep{Lele1992}.
The pressure mesh is half-staggered from the velocity mesh to prevent spurious pressure perturbations~\citep{Laizet2009}.

The simplicity of the Cartesian mesh is leveraged through a robust 2D domain decomposition strategy based on standardized MPI, which imparts excellent strong and weak scaling properties to the code.
The computational domain is divided into ``pencils'', each managed by an MPI process.
Derivatives, interpolations, and one-dimensional FFTs in the x, y, and z directions are performed within the X, Y, and Z pencils, respectively, enabling efficient computation.
Each operation is conducted sequentially in one direction, with global transpositions facilitating the transition from one pencil to another~\citep{Laizet2011}.

\subsubsection{Actuator Disk Model}
To model the effects of the wind turbines in the flow, a number of different methods are available.
Commonly used methods include the actuator surface~\citep{shenActuatorSurfaceModel}, actuator line~\citep{sorensenNumericalModelingWind2002} or actuator disc~\citep{mikkelsenActuatorDiscMethods} models.
This work uses the actuator disc model due to its relatively lower computational cost.
The work of~\citet{revazLargeEddySimulationWind2021} showed that the actuator disc model can provide accurate results of the far wake flow as well as thrust and power predictions.
More advanced methods like the actuator line or surface models may offer better accuracy but are too computationally intensive for this study.

The implementation of the actuator disk model used here is detailed and validated in~\citet{bempedelisTurbulentEntrainmentFinitelength2023}.
It is based on the method of~\citet{calafLESWindTurbine2011} which uses 1D momentum theory and models the wind turbine as a porous disc covering the swept area of the wind turbine blades, $A$.
This imparts a uniform drag force on the flow characterised by,
\begin{equation}\label{eq:adm}
        F_t = -\frac{1}{2} \, \rho \, A \, C_T \left<\frac{U_r}{(1-a)}\right> ^2 \;,
\end{equation}
where $\left<\mathellipsis\right>$ denotes an averaging over the actuator disc and $U_r$ is the rotor normal velocity.
The axial induction factor, $a$, is derived from the thrust coefficient, $C_T$, such that $a= \frac{1}{2}\left(1 - \sqrt{1-C_T}\right)$.
This force is distributed over the area of the disc with a super-Gaussian smoothing applied, as described in~\citet{kingOptimizationWindPlant2017}.
The power output from the turbine can be calculated as,
\begin{equation}\label{eq:admpower}
        P = -F_t U_r \;.
\end{equation}

\subsubsection{Atmospheric Boundary Layer (ABL)}\label{subsubsec:abl}
\begin{figure}[htp]
    \begin{center}
        \includegraphics[width=0.9\linewidth]{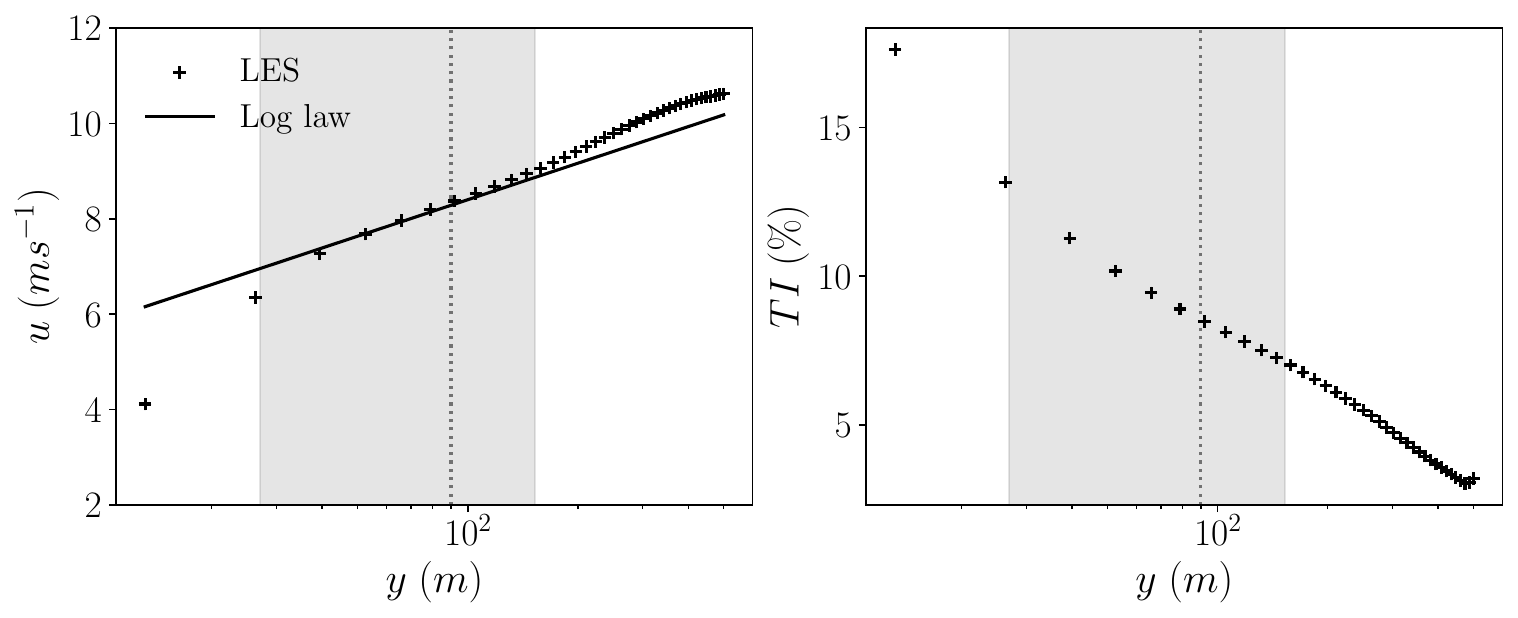}
        \caption{Atmospheric boundary layer profiles for \textit{left: }mean stream-wise velocity and \textit{right: }turbulence intensity.}
        \label{fig:abl}
    \end{center}
\end{figure}

To generate the inflow conditions for the LES wind farm environment, a precursor simulation is conducted to generate an atmospheric boundary layer.
The atmospheric boundary layer is generated using a friction velocity $u^* = 0.442 ms^{-1}$, a boundary layer height $\delta = 504 m$, and a roughness length $z_0 = 0.05 m$.
These parameters were chosen to match those used in previous studies of the Horns Rev wind farm in~\citet{bempedelisDatadrivenOptimisationWind,wuModelingTurbineWakes2015} that best match the wind speed and turbulence intensity conditions observed and reported by~\citet{barthelmieModellingMeasuringFlow2009}.
The profile of the mean stream-wise velocity and the turbulence intensity of the atmospheric boundary layer used for the inlet conditions of the wind farm simulations are shown in Fig.~\ref{fig:abl}.
The location of an NREL-5MW wind turbine within the boundary layer is shown by the shaded region extending across the range of the turbine diameter of $126m$ with the dotted line at the turbine hub height of $90m$.
At this hub height, the atmospheric boundary layer used in the LES experiments has a velocity of $\overline{u_x} = 8.2 ms^{-1}$ and turbulence intensity $TI = 8.6\%$.
These are close to the reported real conditions in~\citet{barthelmieModellingMeasuringFlow2009} of $\overline{u_x} \approx 8 ms^{-1}$ and $TI \approx 8\%$.
At each timestep, a 2D slice of the flow field, orientated normal to the stream-wise direction, is stored, to be used as the inflow condition for the wind farm simulations.

\subsection{Wake Steering}\label{subsec:steering}
\begin{figure}[htp]
    \begin{center}
        \includegraphics[width=0.8\linewidth]{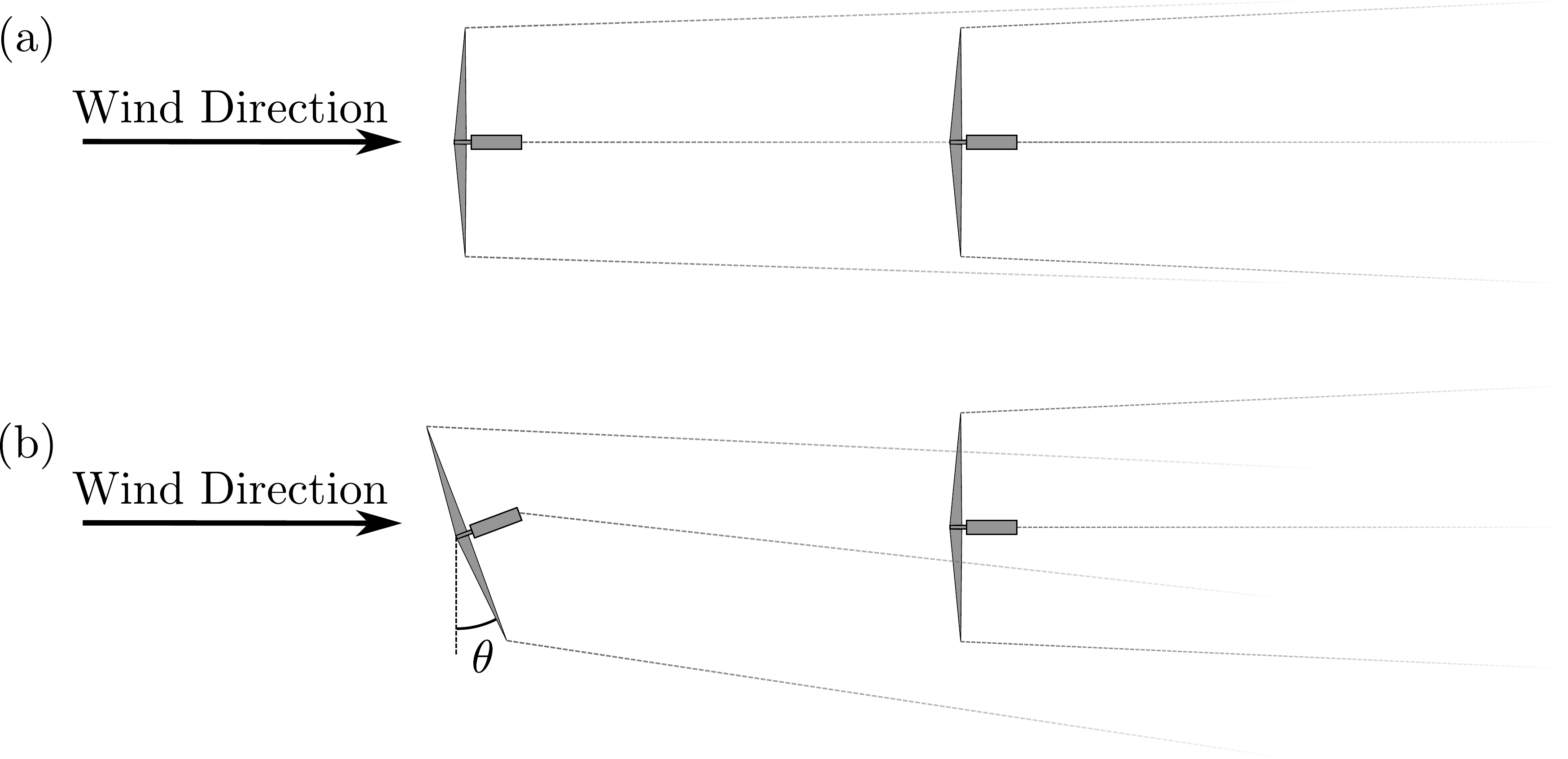}
        \caption{Diagram showing the effect of wake steering showing \textit{(a)} an unsteered case and \textit{(b)} a steered case with the front turbine angled by $\theta$.}
        \label{fig:steer}
    \end{center}
\end{figure}

When performing wake steering in a wind farm setting, the aim is to find the optimum set of wind turbine yaw angles relative to the incoming wind direction that maximise the total power output of the wind farm.
Fig.~\ref{fig:steer} shows a top-down diagram of two turbines aligned with the wind direction in unsteered and steered conditions.
As the total power output of the wind farm is the sum of the power output from the individual wind turbines, this optimisation problem can be expressed as,
\begin{equation}\label{eq:optimise}
        \argmax_{\theta} \sum_{n=1}^N P_n(\theta)
        \quad
\text{subject to}
        \quad
    \theta \in [-\theta_b, \theta_b]
    \;,
\end{equation}
where $\theta$ are the turbine yaw angles and $P_n(\theta)$ are the wind turbine powers for each of the $N$ turbines.
The problem is constrained to a limited range of angles given by $\theta_b$ and in this work $\theta_b = 40^{\circ}$.

\section{Bayesian Optimisation (BO)}\label{sec:bo}

The problem of finding the maximum power output of a wind farm by controlling parameters of the turbines presents an optimisation problem that can only be accurately evaluated at discreet locations in the parameter space.
%using the expensive method of LES\@.
This makes many typical optimisation algorithms (such as ADAM, L-BFGS-B, or differential evolution) that rely on many evaluations or gradients of the function to be known, difficult to perform.
BO presents an efficient optimisation strategy for expensive to evaluate black box functions and is therefore a good candidate for use with LES\@.
BO uses a surrogate model approach to globally iterate towards an optimum solution.

Generally, four steps are required to perform an iteration of a BO~\citep{SnoekPracticalBO2012}.
The first of these is to generate a surrogate model of the quantity of interest with respect to the input parameters.
This is most commonly done using a GP regression model.
Once a surrogate model has been obtained, an acquisition function needs to be calculated.
The third step requires an optimisation of the acquisition function to find the maximum.
This is then used to select the parameters for an additional experiment which is run to augment the current dataset.
This is then iterated until some stopping criterion is met.
In this work, single- and multi-fidelity BO is implemented using the GPyTorch library~\citep{gardner2018gpytorch} and components of the NUBO library~\citep{diessner2023nubo}.

\subsection{Single-Fidelity Bayesian Optimisation} \label{subsec:sfbo}
\begin{figure}[htp]
    \begin{center}
        \includegraphics[width=1\linewidth]{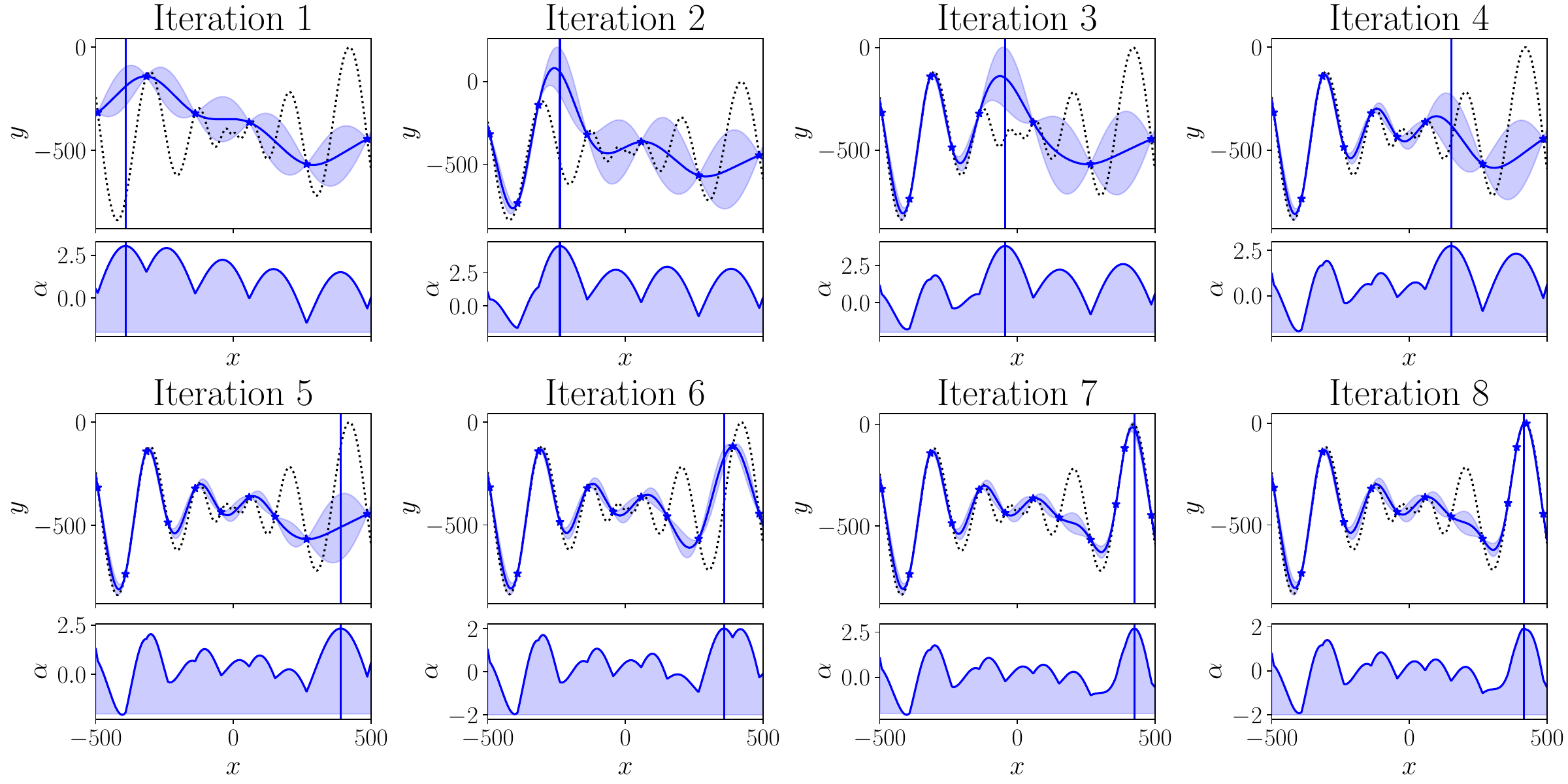}
        \caption{Single-fidelity model, acquisition function and location of the next experiment $\bm{x}_{t+1}$ for 8 BO iterations to find the maximum of the Schwefel function.}
        \label{fig:sfbo}
    \end{center}
\end{figure}

Typically, only a single-fidelity BO is conducted~\citep{bempedelisDatadrivenOptimisationWind,jane-ippelHighfidelitySimulationsWaketowake2023}.
In this work single-fidelity BO is tested against a multi-fidelity approach.
The single-fidelity BO is implemented using GP regression to construct a surrogate model of the quantity of interest over the parameter space.
This is shown for an analytical test function in Fig.~\ref{fig:sfbo}.
The test function used is the one dimensional Schwefel function and is shown by the dotted line.
This function was used as it is a particularly challenging function with many local optima and is commonly used for testing optimisation frameworks~\citep{yang2010engineering}.
For each of eight iterations, the updated model is shown.
This surrogate model is then used to construct an upper confidence bound (UCB) acquisition function in the parameter space that is then maximised to determine the optimal location of the next evaluation before repeating the process.
The details of the GP model and the UCB acquisition function are given below.

\subsubsection{Gaussian Process (GP)}
GP models are constructed based on a set of data, $\mathcal{D} = \left\{ (\bm{x}_1, y_1), \dots, (\bm{x}_n, y_n) \right\}$, where $(\bm{x}, \bm{y}) \in \mathbb{R}^{n \times d} \times \mathbb{R}^n$ are the $n$ number of inputs and outputs from our black-box function respectively.
To learn a model of the form $\bm{y} = f(\bm{x})$, a zero-mean GP prior is placed on $f(\bm{x})$, such that $f(\bm{x}) \sim \mathcal{GP}(0, k(\bm{x}, \bm{x}'))$, where the covariance function $k(\bm{x}, \bm{x}')$ is a kernel function that quantifies the similarity between points $\bm{x}$ and $\bm{x}'$.

To obtain a posterior distribution, the joint Gaussian prior distribution is conditioned on the observations $\mathcal{D}$.
This predictive posterior distribution $p(f_*(\bm{x}_*) | \mathcal{D})$ at a test point $(\bm{x}_*)$ is also Gaussian, with a mean $\mu(\bm{x}_*)$ and variance $\sigma^2(\bm{x}_*)$ that can be calculated as:
\begin{equation}\label{eq:gpr_mu}
\mu(\bm{x}_*) = \bm{k}_x \bm{K}^{-1} \bm{y}
\;,
\end{equation}
\begin{equation}\label{eq:gpr_sigma}
\sigma^2(\bm{x}_*) = k(\bm{x}_*, \bm{x}_*) - \bm{k}_x \bm{K}^{-1} \bm{k}_x^T
\;,
\end{equation}
where $k_x = [k(\bm{x}_*,\bm{x}_1), \dots, k(\bm{x}_*,\bm{x}_n)]$ is the covariance vector between the new input point and the observed inputs and $\bm{K}$ is the $n \times n$ covariance matrix with entries $K_{ij} = k(x_i, x_j)$ for $i, j = 1, \ldots, n$.
The posterior mean $\mu$ can then be used to predict the quantity of interest at any point in the parameter space with the posterior variance providing a measure of the uncertainty.
The posterior mean $\mu$ and standard deviation $\sigma$ are shown in Fig.~\ref{fig:sfbo} by the solid line and shaded region respectively for each iteration of the BO.
Further details on the implementation of GP regression can be found in~\citet{rasmussenGaussianProcessesMachine2008}.

\subsubsection{Upper Confidence Bound (UCB) Acquisition Function}

Within the BO loop, at each iteration $t$, the data $\mathcal{D}$ is expanded by selecting a new set of inputs $\bm{x}_t$ and evaluating them to find $\bm{y}_t$.
To find the next set of input parameters $\bm{x}_{t+1}$ an acquisition function is used.
A commonly used acquisition function, adopted in this work, is the UCB acquisition function~\citep{auer2002finite,srinivas2009gaussian}.
It combines the expected value from the mean $\mu_t(\bm{x})$, and of the uncertainty $\sigma_t(\bm{x})$ captured by the standard deviation of the GP model.
These terms are weighted with the term $\beta_t$ to provide a tradeoff between exploration of the parameter space and the exploitation of the current expected maximum.
The UCB acquisition function is given by,
\begin{equation}\label{eq:ucb}
    \alpha^{UCB}_t(\bm{x}) = \mu_t(\bm{x}) + \beta_t \sigma_t(\bm{x}) \;.
\end{equation}

The location of the next experiment in the parameter space is then found by finding the maximum location of the acquisition function,
\begin{equation}\label{eq:ucbopt}
    \bm{x}_{t+1} = \argmax_x\left(\alpha^{UCB}_t(\bm{x})\right) \;.
\end{equation}
This can be done using gradient-based optimisation methodologies, and in this work the L-BFGS-B~\cite{liuLimitedMemoryBFGS1989} optimiser is used for its efficiency.
Fig.~\ref{fig:sfbo} shows the acquisition function and location of the next experiment $\bm{x}_{t+1}$ determined at its maximum for each iteration of the optimisation.

\subsection{Multi-Fidelity Bayesian Optimisation}\label{subsec:mfbo}

In the current work, a multi-fidelity BO strategy is implemented to exploit the use of cheap low-fidelity approximations in the form of analytical wake models, whilst also benefiting from the additional physics captured by the high-fidelity LES\@.
To combine multiple fidelities in a BO, the model and acquisition function require modification.
In this work, a multi-fidelity GP model is constructed in the form of a Multi-Fidelity NARGP~\citep{Perdikaris2017}.
A multi-fidelity acquisition function is constructed, based on the UCB acquisition function, to provide the location for the next experiment as well as the fidelity that should be used.

\subsubsection{Multi-Fidelity Non-Linear Auto-Regressive Gaussian Process (NARGP)}
\begin{figure}[htp]
    \begin{center}
        \includegraphics[width=0.6\linewidth]{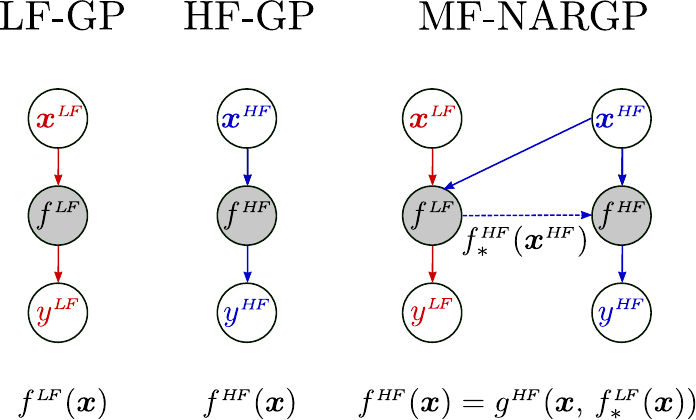}
        \caption{Diagram showing the structure of the multi-fidelity NARGP model compared to the standard structure of the low- and high-fidelity GP models.}
        \label{fig:nargp}
    \end{center}
\end{figure}

The multi-fidelity BO is implemented by constructing a multi-fidelity surrogate model using the NARGP approach~\citep{Perdikaris2017}, the structure of which is displayed in Figure~\ref{fig:nargp}.
This method is a generalisation of the auto-regressive scheme of~\citet{kennedyBayesianCalibrationComputer2001} that allows for complex nonlinear relationships between the model fidelities to be captured.
Assuming a range of different fidelity datasets exist $\mathcal{D}^m = (\bm{x}^m, \bm{y}^m)$ with $m = 1, \dots, M$ where $\mathcal{D}^1$ is the lowest fidelity dataset ranging to $\mathcal{D}^M$ as the highest fidelity dataset.
The lowest fidelity dataset is used to form a GP model, $f^1$ as in Sect.~\ref{subsec:sfbo}.
Subsequent fidelity models can then be formed as,
\begin{equation}\label{eq:nargp}
f^{m+1}(\bm{x}) = g^{m+1}(\bm{x}, \, f^m_*(\bm{x})) \;.
\end{equation}

\begin{figure}[htp]
    \begin{center}
        \includegraphics[width=0.7\linewidth]{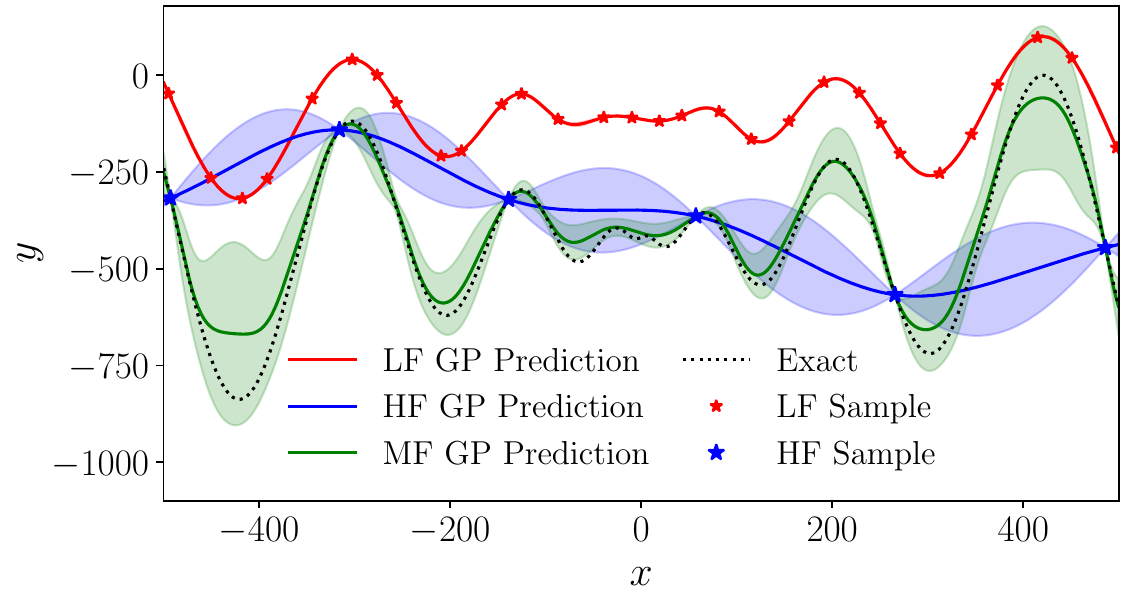}
        \caption{Comparison between a low-fidelity GP model, high-fidelity GP model and a multi-fidelity NARGP model on a one dimensional Schwefel function. The shaded regions show the confidence intervals of the model predictions.}
        \label{fig:mf_analytical}
    \end{center}
\end{figure}

The posterior prediction from the previous fidelity, $f^m_*(\bm{x})$, forms an additional input features to the subsequent model, $f^{m+1}$, and it is this inclusion into the model that allows the bridging between the two fidelities.
In this work, $M=2$ fidelities are used with $\mathcal{D}^1 = \mathcal{D}^{LF}$ and $\mathcal{D}^2 = \mathcal{D}^{HF}$.
The structure of a low-fidelity GP, high-fidelity GP, and multi-fidelity NARGP with $M=2$ are shown in Fig.~\ref{fig:nargp}.
The models generated with a low-fidelity GP, high-fidelity GP, and multi-fidelity NARGP are shown for a test analytical function in Fig.~\ref{fig:mf_analytical}.
In this case, the $\mathcal{D}^{LF}$ and $\mathcal{D}^{HF}$ datasets include 30 and 6 datapoints respectively.
It can be seen that the high-fidelity model (trained only on $\mathcal{D}^{HF}$) matches the exact solution function at the data points, however, in other regions the model deviates from the exact solution considerably.
Conversely, the low-fidelity model (trained only on $\mathcal{D}^{LF}$) picks up the correct trend across $x$ but, due to the inaccuracies in the data, does not provide an accurate model of the exact function.
When combining both datasets $\mathcal{D}^{LF}$ and $\mathcal{D}^{HF}$ using the multi-fidelity NARGP, the model is close to the exact solution across much of the parameter space, even away from high-fidelity data points.

\subsubsection{Multi-fidelity Acquisition Function}

In the multi-fidelity BO loop, for iteration $t$, once a multi-fidelity surrogate model has been constructed, it can then be used to determine the parameters, $\bm{x}_{t+1}$, and fidelity, $m_{t+1}$, of a subsequent experiment.
A multi-fidelity acquisition function needs to be defined to inform the selection of $\bm{x}_{t+1}$ and $m_{t+1}$ to learn more information about the optimum solution.
In this work, a multi-fidelity acquisition function is constructed, based on the UCB acquisition function in equation~\ref{eq:ucb}.
The ability of the UCB acquisition function to explicitly define the exploration-exploitation trade-off enables the cost ratio between the fidelities to be explicitly defined.
This multi-fidelity UCB (MFUCB) acquisition function uses the mean of the highest-fidelity part of the surrogate model but includes the standard deviation from all fidelities.
These are combined to give,
\begin{equation}\label{eq:mfucb}
    \alpha^{MFUCB}_t(\bm{x}) = \max_{m}\left(\mu^{M}_t(\bm{x}) + \beta^{m}_t \sigma^{m}_t(\bm{x})\right) \;,
\end{equation}
where $\beta^{m}$ are the exploration/exploitation tradeoff parameters for each fidelity.
These can be related by the cost ratio between the fidelities so that, $\beta_t^{m} = \gamma_t^{m}\beta_t^{M}$.
By finding the maximum of this equation~\ref{eq:mfucb}, both the location in the parameter space and the appropriate fidelity can be selected:
\begin{equation}
    (\bm{x}_{t+1}, \;\; m_{t+1}) = \argmax_{x,m}\left(\alpha^{MFUCB}_t(\bm{x})\right) \;.
\end{equation}

\begin{figure}[htp]
    \begin{center}
        \includegraphics[width=1\linewidth]{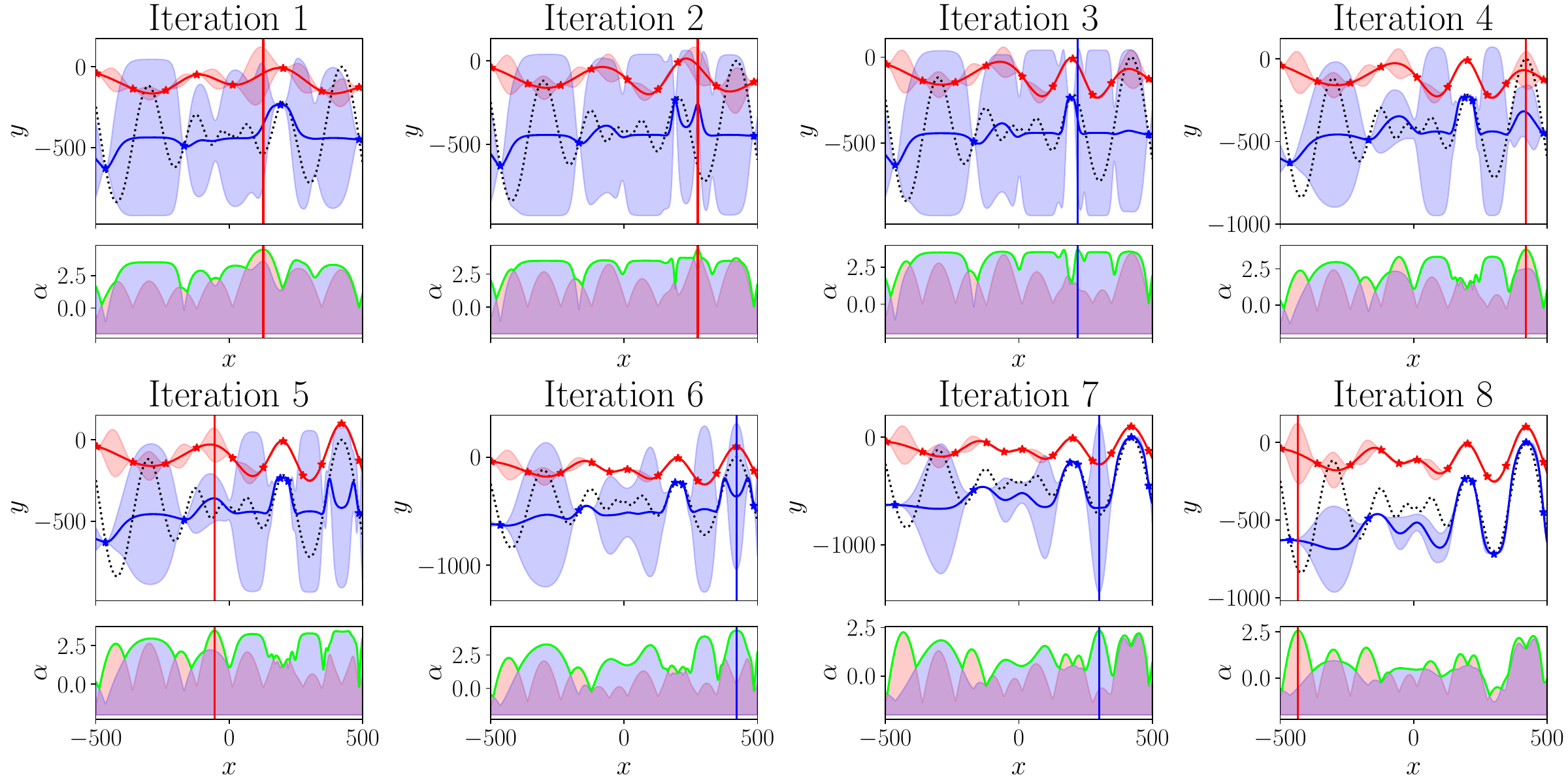}
        \caption{Multi-fidelity model, acquisition function and location and fidelity of next experiment $(\bm{x}_{t+1}, \;\; m_{t+1})$ for 8 BO iterations to find the maximum of the Schwefel function.}
        \label{fig:mfacq}
    \end{center}
\end{figure}

The form of the acquisition function for an analytical test function with $M=2$ is shown in Fig.~\ref{fig:mfacq}.
The top plots show the low- and high-fidelity parts of the multi-fidelity model trained on the respective datasets compared to the exact function.
The lower plots show the MFUCB acquisition function and its constitutive low- and high-fidelity components shaded.
The vertical line shows the location in parameter space and the fidelity of the next experiment $(\bm{x}_{t+1}, \;\; m_{t+1})$.
At iteration 1, 2, 4, 5 and 8, the acquisition function defines the fidelity of the next experiment to be low-fidelity and the model is updated accordingly in the next iteration.
Iterations 3, 6 and 7 define the next experiments as high-fidelity.
This approach allows for the optimum to be found with fewer high-fidelity samples.

\subsection{Stopping Criterion}

In the current work, the optimisation is stopped when all of the parameters in two consecutive iterations are within a set tolerance $\lambda$ so that,
\begin{equation}
\| \bm{x}_{t} - \bm{x}_{t-1} \| / N < \lambda
\;.
\end{equation}
Once this condition is met, the optimisation is stopped.
When multiple fidelities are used, only iterations using the highest fidelity $M$ are included in the evaluation of the convergence criteria.

\section{Results}\label{sec:results}

\subsection{Single Turbine Control}\label{subsec:1turbine}

A relatively simple wind farm layout of two aligned wind turbines is used to provide a proof of concept for the multi-fidelity optimisation approach.
The wind farm set-up consists of two NREL 5MW turbines, each with a diameter of $D=126m$ and with a hub-height of $90m$.
The turbines are separated in the direction of the freestream wind by a distance of $7D$.
The wind conditions, detailed in section~\ref{subsubsec:abl}, are used with a velocity of $\overline{u_x} = 8.2 ms^{-1}$ and turbulence intensity $TI = 8.6\%$ at the turbine hub height.
The yaw angle of the downstream turbine is fixed to $\theta_2 = 0^{\circ}$, to align with the free stream wind direction, and the upstream turbine yaw angle ($\theta_1$) is optimised for maximum wind farm power output.
The second wind turbine is not controlled as its wake will not affect downstream turbines.
Although it is not expected that significant power improvement will be found for this wind farm layout due to the limited number of turbines, the optimisation being one dimensional allows for the methods described in section~\ref{sec:bo} to be clearly analysed.
Initial datasets of yaw angles are generated using latin hypercube sampling~\citep{mckayComparisonThreeMethods1979} to obtain the yaw angle $\theta_1$.
This approach is used to randomly select the sampling locations whilst ensuring that they represent the variability within the parameter space.

\subsubsection{Low-Fidelity Optimisation}\label{subsubsec:lf}

\begin{figure}[htp]
    \begin{center}
        \includegraphics[width=0.33\linewidth]{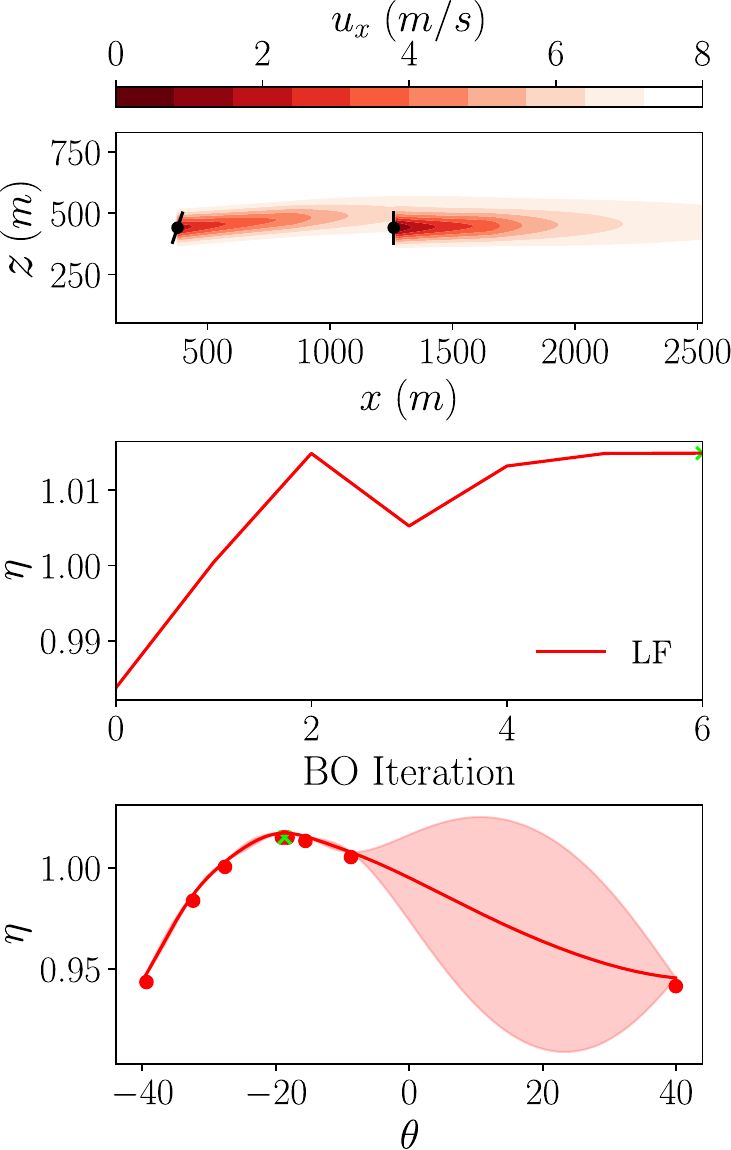}\hfill
        \includegraphics[width=0.33\linewidth]{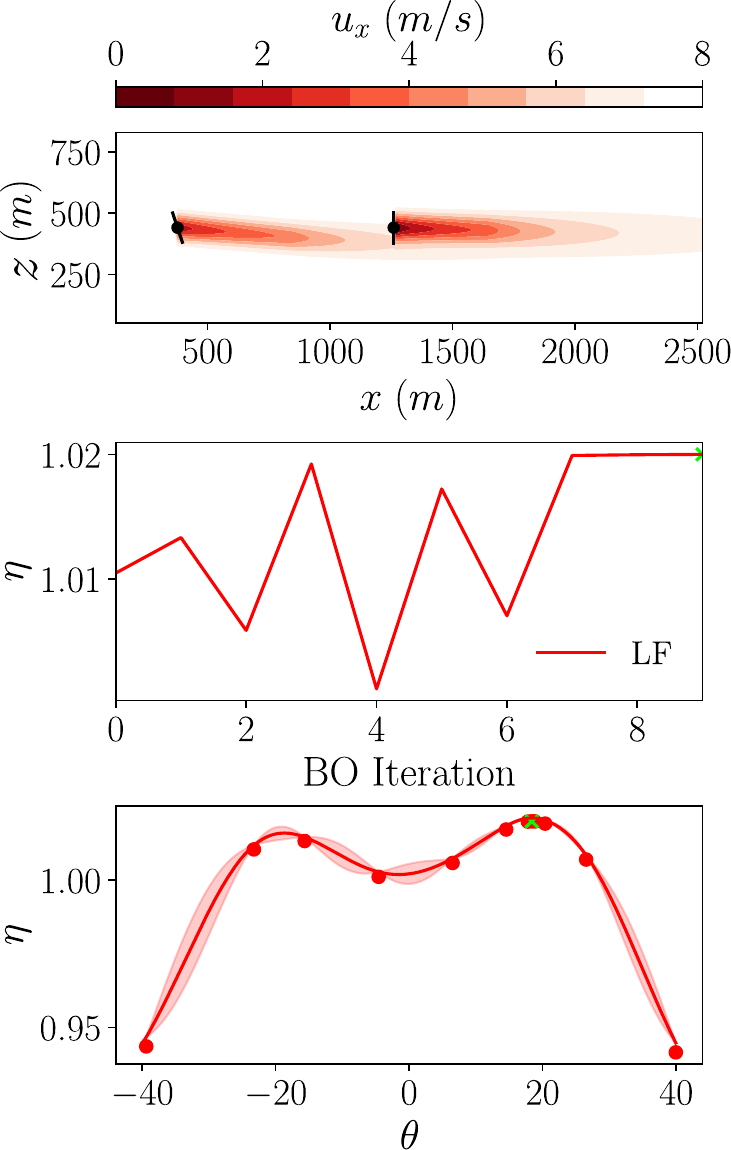}\hfill
        \includegraphics[width=0.33\linewidth]{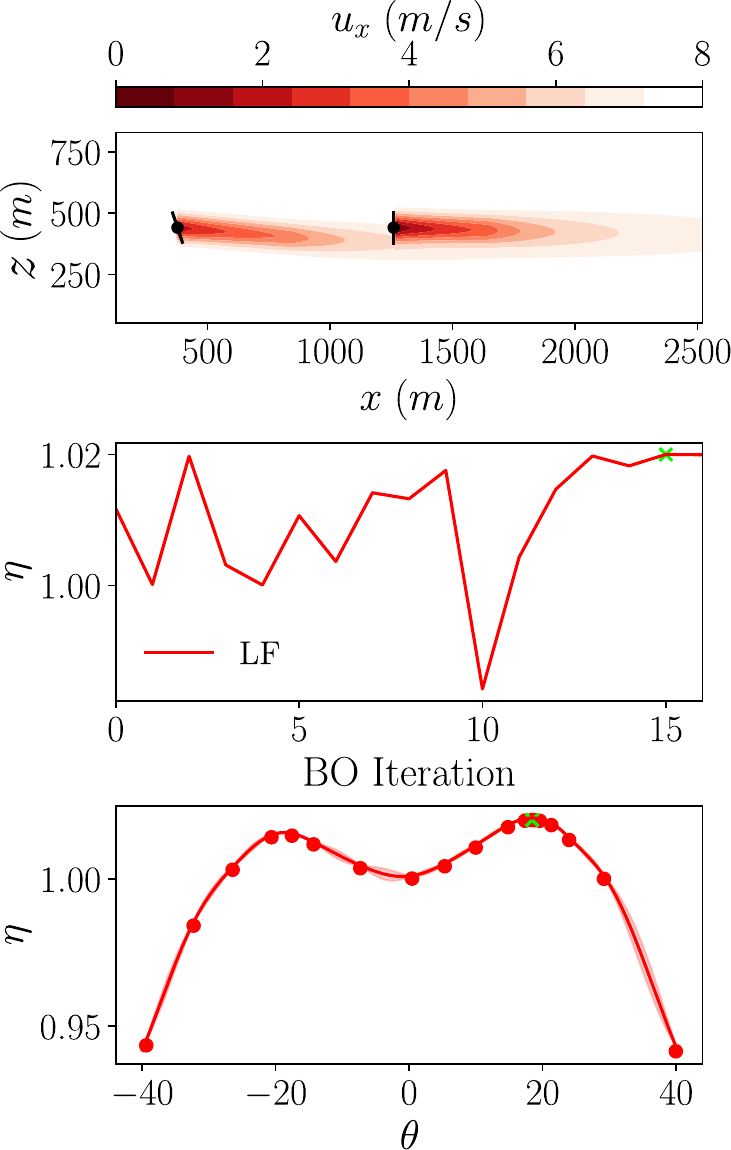}
        \caption{Low-fidelity BO results for yaw of the front turbine in a 2x1 turbine array with different values for $\beta$ (\textit{left: }$\beta=1$ \textit{center: }$\beta=4$ \textit{right: }$\beta=16$). \textit{Top: }Slice of stream-wise velocity at hub height. \textit{Center: }Total wind farm efficiency calculated at each BO iteration. \textit{Bottom: } Low-fidelity GP model at the final iteration.}
        \label{fig:1t-lf}
    \end{center}
\end{figure}

Firstly, a single-fidelity BO is performed using low-fidelity data, and we will refer to this as a low-fidelity BO\@.
The initial dataset consists of a set of three yaw angles, and for each configuration the corresponding total power output of the wind farm is calculated using the GCH wake model.
The BO is performed using three different exploration/exploitation parameters from equation~\ref{eq:ucb}, with $\beta=1$, $\beta=4$ and $\beta=16$.
This will show the effect of the parameter on the final surrogate model and the number of iterations required to reach an optimum result.
In each case, the optimisation is stopped once the stopping criterion is met with $\lambda =0.5^{\circ}$.
The results of these low-fidelity BOs are shown in Fig.~\ref{fig:1t-lf}.
For each case, the top plot shows a slice of the velocity field at the turbine hub height, the second plot shows the power output obtained from the GCH wake model at each BO iteration, and the final plot shows the final low-fidelity GP model obtained from the optimisation.
In each case, the total power output from the wind farm is normalised by the total power output of unoptimised layout with $\theta_i = 0$ evaluated in the highest fidelity used in that optimisation.
This can be expressed as:
\begin{equation}\label{eq:efficiency}
        \eta = \frac{ \sum_{n=1}^N P^m_n(\theta)}{\sum_{n=1}^N P^M_n(\theta=0^{\circ})} \;.
\end{equation}

With $\beta = 1$, the optimum angle found is $\theta_1 = -18.7^{\circ}$ producing a power improvement of $1.5\%$ over an uncontrolled case ($\theta_1 = 0^{\circ}$).
However, with $\beta = 4$ and $\beta = 16$, the optimum angle found is $\theta_1 = 18.4^{\circ}$ producing a power improvement of $2.0\%$.
This discrepancy between the optimisation results is due to the over-exploitation in the $\beta=1$ case.
The specifics of the initial sampling lead to the local optimum being found without exploring other regions of the parameter space.
In line with this, the confidence intervals of the final surrogate model are large for the $\beta = 1$ case.
As $\beta$ is increased, both local optima are represented in the surrogate model and the confidence intervals are reduced.
As $\beta$ is increased, there is also an increase in the number of BO iterations required to reach a converged solution.
Increasing the $\beta$ parameter from $\beta=1$ to $\beta=4$, and further to $\beta=16$ results in a corresponding rise in the required number of iterations, from 7 to 10 and 17, respectively.
There is therefore a balance to be found between the number of optimisation iterations and the exploration of the parameter space that can be controlled using the $\beta$ parameter.
As, when using the GCH analytical wake model, the computational time spent running an individual experiment for this case is $~1.5 ms$ on a modern workstation (Intel Core i9-13900K), it is worth being more conservative and running with a higher $\beta$ to ensure that the parameter space is sufficiently explored.

\subsubsection{High-Fidelity Optimisation}\label{subsubsec:hf}

\begin{figure}[htp]
    \begin{center}
        \includegraphics[width=0.33\linewidth]{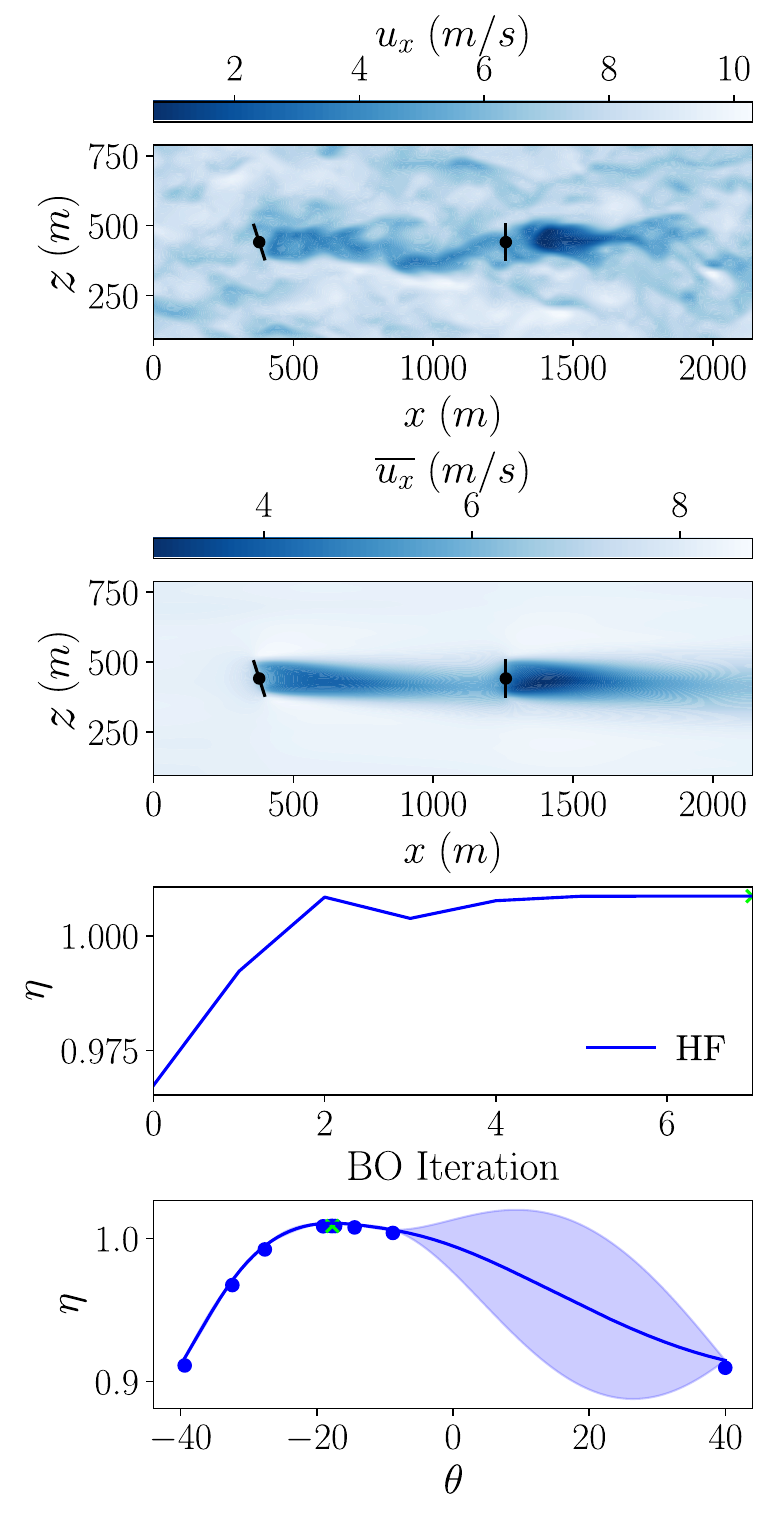}\hfill
        \includegraphics[width=0.33\linewidth]{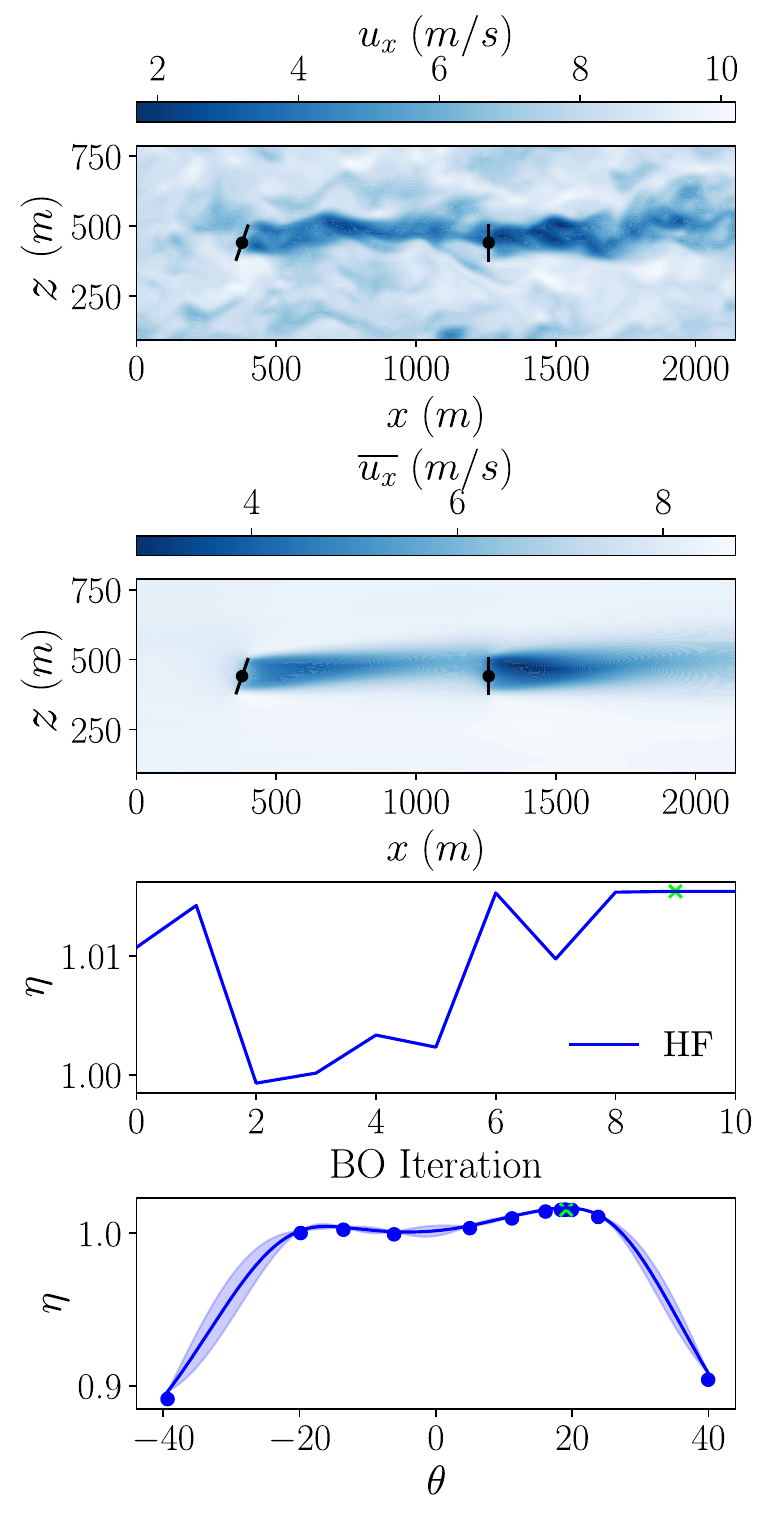}\hfill
        \includegraphics[width=0.33\linewidth]{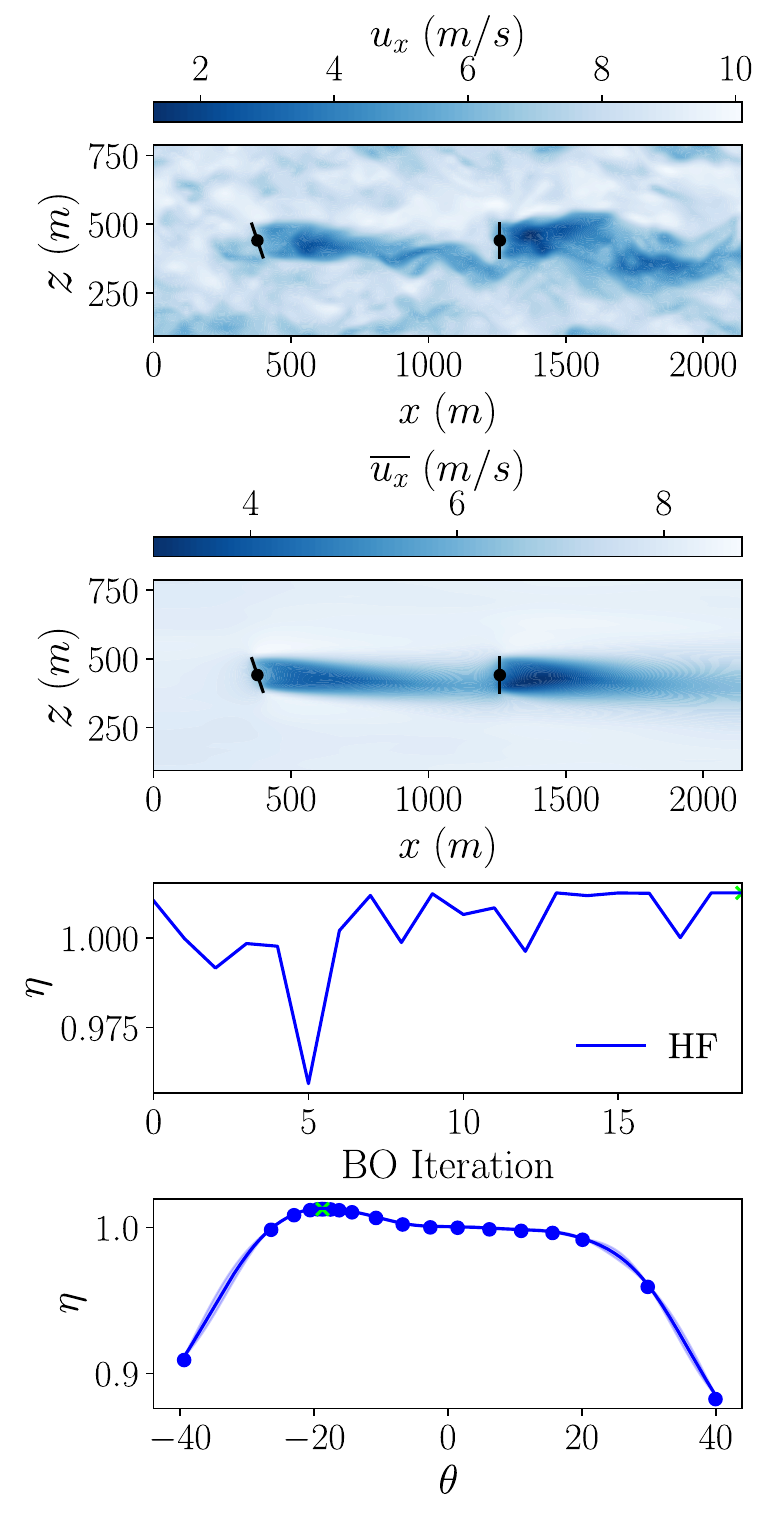}
        \caption{High-fidelity BO results for yaw of the front turbine in a 2x1 turbine array with different values for $\beta$ (\textit{left: }$\beta=1$ \textit{center: }$\beta=4$ \textit{right: }$\beta=16$). \textit{Top: }Slice of stream-wise velocity at hub height from LES. \textit{Center: }Total wind farm efficiency calculated at each BO iteration. \textit{Bottom: } High-fidelity GP model at the final iteration.}
        \label{fig:1t-hf}
    \end{center}
\end{figure}

For comparison, a single-fidelity BO based on high-fidelity LES data is also conducted on this wind farm set-up.
In each case, the initial dataset consists of a set of 2 yaw angles and the corresponding total power output of the wind farm calculated using LES and the ADM\@.
The LES is performed on a uniform grid with dimensions $L_x \times L_y \times L_z = 17D \times 500m \times 7D$ and a grid spacing of $D/10$ in the three spatial directions.
The power output from the LES calculation is averaged over a simulated time period of $\approx 2$ hours with a timestep of $0.2s$.
As above, the BO is performed using three different exploration/exploitation parameters from equation~\ref{eq:ucb}, with $\beta=1$, $\beta=4$ and $\beta=16$ and the optimisation is stopped when the stopping criterion with $\lambda = 0.5^{\circ}$ is met.
The results of these high-fidelity BOs are shown in Fig.~\ref{fig:1t-hf}.
Plots of the instantaneous and time averaged velocity fields, the power output obtained at each BO iteration, and the final high-fidelity GP model obtained from the optimisation are shown for each case.

Again, as $\beta$ is increased, there is a corresponding increase in the number of BO iterations required to reach a solution.
With $\beta = 1$, the optimum angle found is $\theta_1 = -17.7^{\circ}$ producing a power improvement of $0.9\%$ over an uncontrolled case ($\theta_1 = 0^{\circ}$).
With $\beta = 4$, the optimum angle found is $\theta_1 = 19.1^{\circ}$ producing a power improvement of $1.5\%$ over an uncontrolled case ($\theta_1 = 0^{\circ}$).
With $\beta = 16$, the optimum angle found is $\theta_1 = -18.7^{\circ}$ producing a power improvement of $1.3\%$ over an uncontrolled case ($\theta_1 = 0^{\circ}$).

The power improvement from the high-fidelity optimisation, calculated from the LES, is lower than the power improvement found with the low-fidelity wake model at $1.5\%$ compared to $2.0\%$.
The optimum yaw angles obtained in the low- and high-fidelity optimisations are within $1^{\circ}$ of each other.
This suggests that the GCH wake model used in the low-fidelity optimisation works well for this two-turbine case and that much of the involved physics is captured in the model.
The discrepancy in the results may come from the lack of rotation effects modelled by the actuator disc model in the LES\@.
When comparing the final model generated by the GPs in the low- and high-fidelity cases, the low-fidelity model shows a larger drop-off in the power at high-yaw angles than the high-fidelity model.
There is also some variation in the results for the high-fidelity model between cases.
This likely is due to the stochastic nature of the LES and the large variations in the flow conditions due to the turbulent nature of the atmospheric boundary layer, as can be seen in the instantaneous flow fields at the top of Fig~\ref{fig:1t-hf}.

The computational cost of running the LES is significantly greater than the analytical wake model.
Each new experiment took over $12$ minutes to run on 126 cores (dual AMD EPYC 7742 processors), making the LES $~60\times10^6$ times more computationally expensive than running the wake model.
It is therefore much more important to limit the number of experiments conducted.
This could be done in part by selecting an appropriate exploration parameter, $\beta$ however, this is difficult to select a priori without under-exploring the parameter space and finding a local optimum rather than the global optimum.
Applying the multi-fidelity optimisation approach should allow the number of LES runs to be limited whist still sufficiently exploring the parameter space.

\subsubsection{Multi-Fidelity Optimisation}\label{subsubsec:mf}

% Gauss - GCH
In this section, the multi-fidelity optimisation approach, described in section~\ref{subsec:mfbo}, is applied to the yaw control of a single turbine.
In all cases, initial datasets are generated using latin hypercube sampling with 6 low-fidelity and 3 high-fidelity yaw configurations and optimisation coefficients of $\beta_{HF} = 1$ and $\gamma_t^{LF} = 8$ are used.
These optimisation coefficients correspond to a well explored parameter space in the low-fidelity and an under-explored parameter space in the high-fidelity, as found in section~\ref{subsubsec:lf} and~\ref{subsubsec:hf}.

\begin{figure}[htp]
    \begin{center}
        \includegraphics[width=0.32\linewidth]{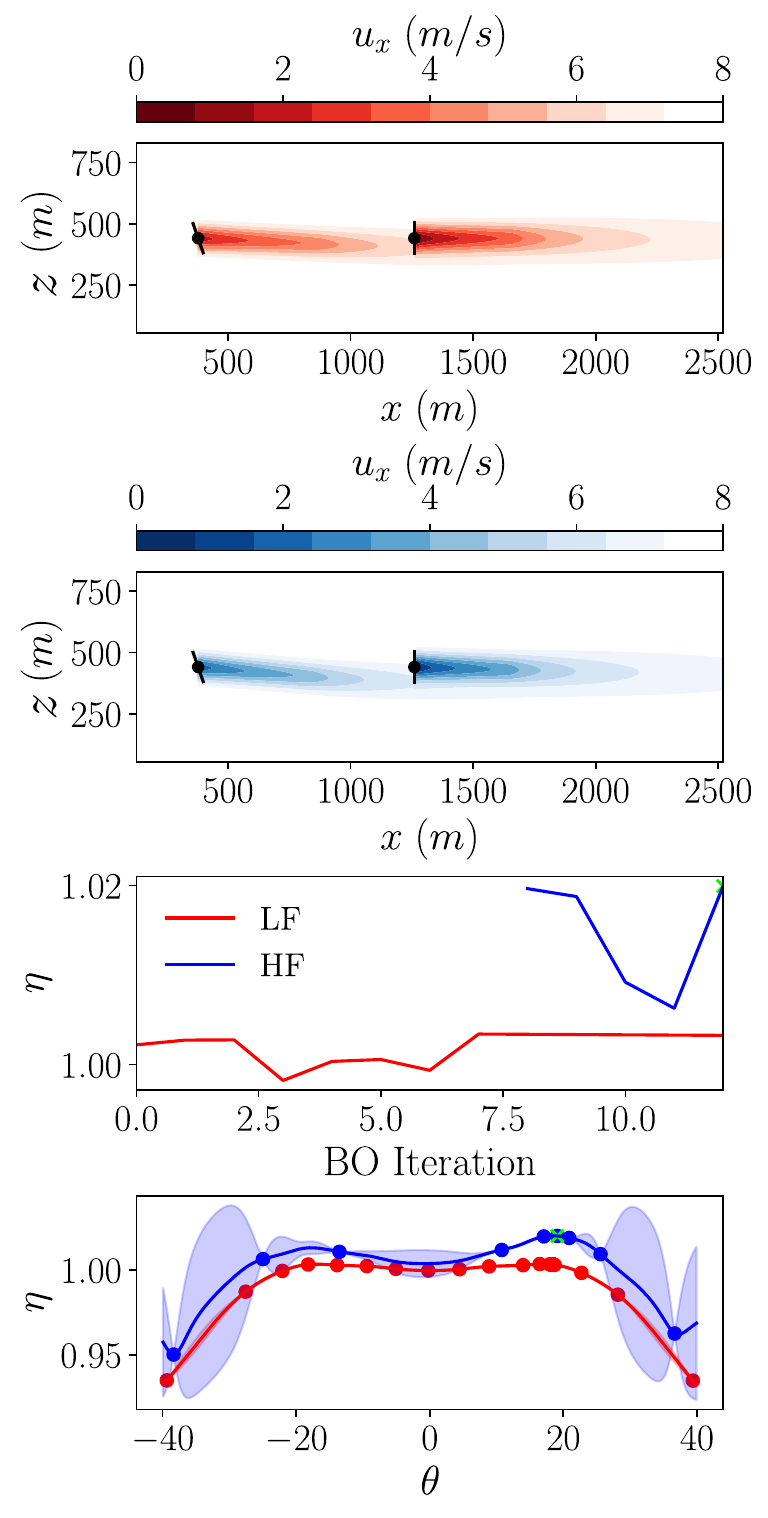}
        \includegraphics[width=0.32\linewidth]{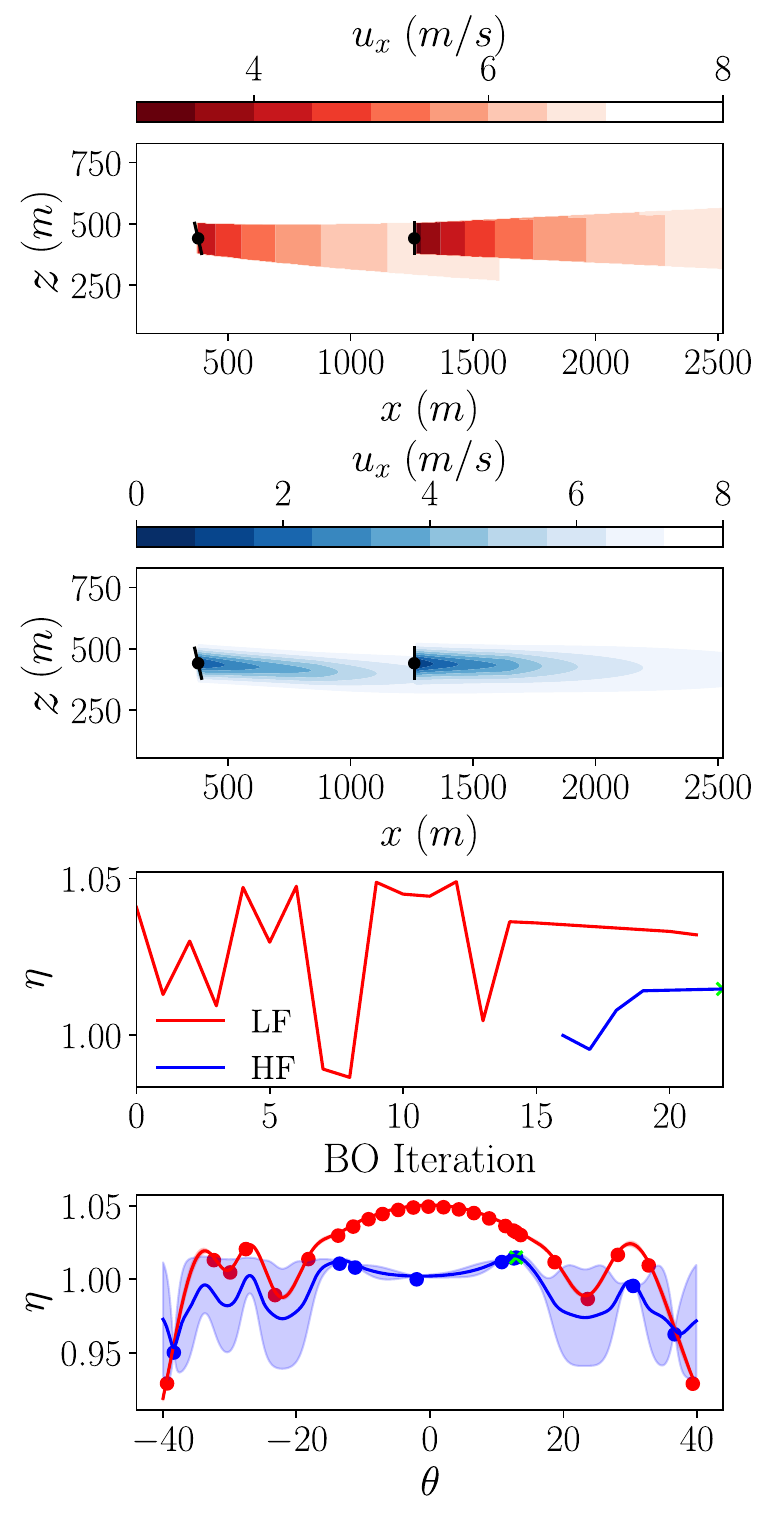}
        \includegraphics[width=0.32\linewidth]{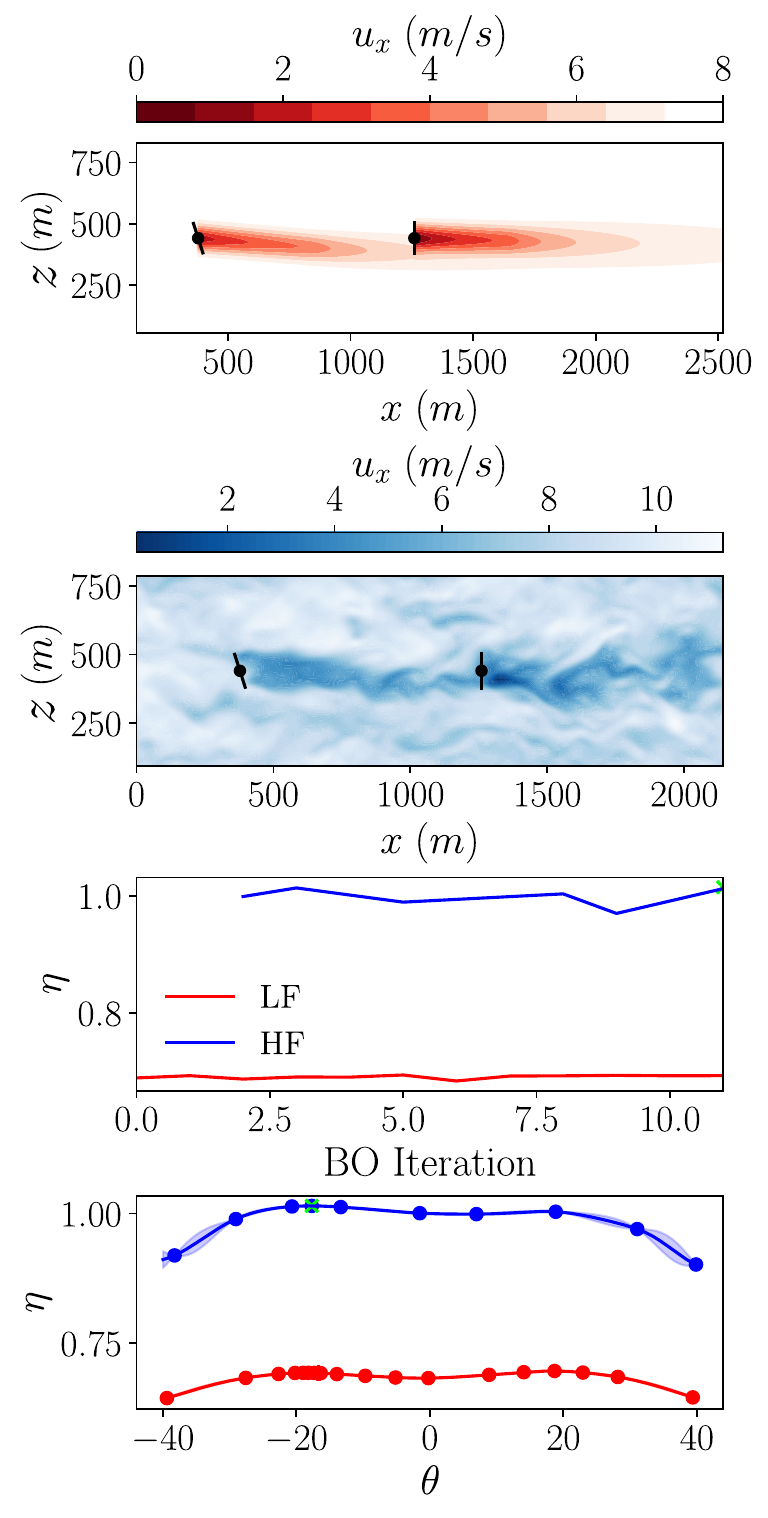}
        \caption{Multi-fidelity BO result for yaw of the front turbine in a 2x1 turbine array using \textit{left: }Gaussian-GCH \textit{center: }Jensen-GCH \textit{right: }GCH-LES fidelities.}
        \label{fig:1t-mf}
    \end{center}
\end{figure}

The multi-fidelity optimisation is first done with data for both fidelities collected using analytical wake models.
The Gaussian wake model is used to provide the low-fidelity data, and the GCH model is used for the high-fidelity data.
The results of this optimisation are shown in the plots on the left of Fig.~\ref{fig:1t-mf}.
Slices of the flow field at the turbine hub-height are shown for the optimal yaw configuration evaluated with both low-fidelity (in red) and high-fidelity methods (in blue).
The optimum solution of $\theta_1 = 19.0^{\circ}$ is found in 13 optimisation iterations, producing a power improvement of $2\%$.
For the first 8 iterations, the multi-fidelity acquisition function results in low-fidelity experiments being conducted.
The next 5 iterations are then conducted using the high-fidelity.
This finds the same optimisation result as the single-fidelity in Fig~\ref{fig:1t-lf} with a high exploration-parameter, but uses fewer of the high-fidelity evaluations than the case with $\beta=1$.

% Jensen - GCH
The choice of low-fidelity environment is important when conducting the multi-fidelity optimisation.
A second multi-fidelity optimisation is conducted on this case with the same high-fidelity environment of the GCH wake model but with the Jensen wake model used for the low-fidelity.
The Jensen wake model was designed for use in micro-sitting problems and does not perform well for wake steering optimisation due to its uniform velocity deficit distribution in the span-wise direction.
The results, displayed in the centre of Fig.~\ref{fig:1t-mf}, show that more low-fidelity iterations are initially required to capture the more complex response of the power across the parameter space.
This is then followed by 5 high-fidelity iterations before meeting the convergence criteria.
The resulting solution is a yaw angle of $\theta_1 = 12.9^{\circ}$ resulting in a power improvement of $1.5\%$, which is not as good as the optimum found in previous cases.
In this case, the low-fidelity data does not contain useful information for improving the high-fidelity part of the NARGP model and so hinders the proper exploration of the parameter space.
For this reason, the GCH wake model is used to provide the low-fidelity data in the multi-fidelity optimisation with the LES as the high-fidelity data source.

% GCH - LES
A final multi-fidelity optimisation is run on this case using the GCH wake model for the low-fidelity data and LES for the high-fidelity data.
The results of this optimisation are shown on the right of Fig~\ref{fig:1t-mf}.
In this case, there is a much larger offset between the power output from the high- and low-fidelity models.
The optimum angle found is $\theta_1 = -17.7^{\circ}$ producing a power improvement of $1.5\%$ over an uncontrolled case ($\theta_1 = 0^{\circ}$).
This is comparable to the best optimum found using the single-fidelity LES optimisation shown in Fig.~\ref{fig:1t-hf} whilst only using 7 compared to 20 high-fidelity experiments.
This reduction in high-fidelity experiments whilst still finding the same optimum is the equivalent to a saving of more than 330 CPU hours (on AMD EPYC 7742 processors).

Although, in the two-turbine case presented in this section, the GCH model results perform well in comparison to the LES results for finding the optimum yaw angle, there is a difference in the resulting power output calculated.
By using a multi-fidelity optimisation, optimisation results were found that were equivalent to the best high-fidelity optimisation results whilst limiting the cost required to explore the parameter space.
This is expected to be of increased benefit when applied to cases where there is more of a discrepancy between the LES and wake modelling.

\subsection{Horns Rev Wind Farm}\label{subsec:hornrev}

\begin{figure}[htp]
    \begin{center}
        \includegraphics[width=0.65\linewidth]{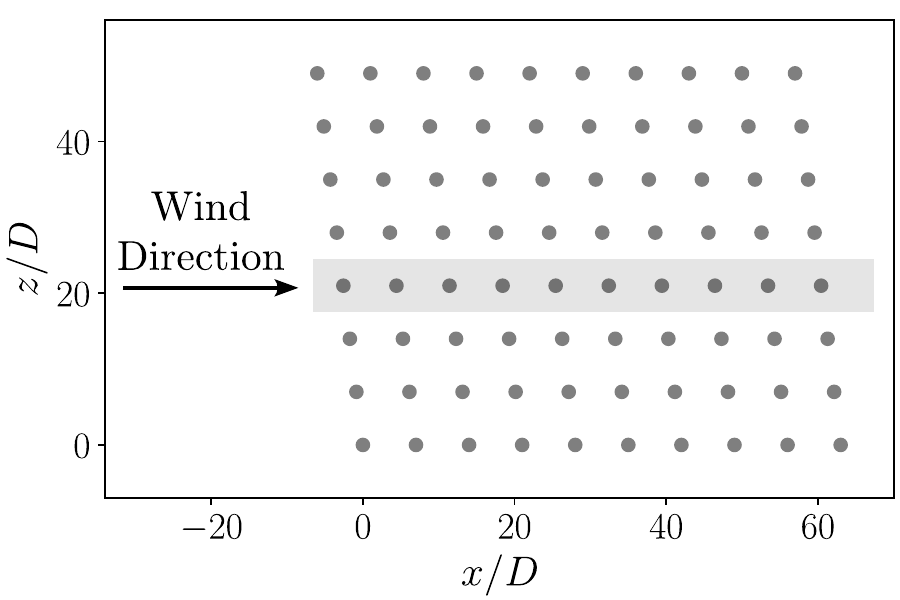}
        \caption{The Horns Rev wind farm turbine locations with LES domain highlighted.}
        \label{fig:hornsrev}
    \end{center}
\end{figure}

The work of~\cite{bempedelisDatadrivenOptimisationWind} showed that, for wake steering applications in the Horns Rev wind farm, an LES-based optimisation could discover yaw angles that provided a higher total wind farm power output than a wake model based optimisation.
In this section, we test the multi-fidelity optimisation approach on a wind farm set-up that is representative of the Horns Rev wind farm.
This wind farm is situated in the North Sea off the coast of Denmark and consists of 80 turbines in an 8 by 10 grid and a spacing between each turbine of $7D$.
The locations of the turbines within the wind farm are shown in Fig.~\ref{fig:hornsrev} with the same turbine parameters used as in section~\ref{subsec:1turbine}.
A westerly wind direction is considered such that the wind is aligned with the turbine rows with a velocity of $\overline{u_x} = 8.2 ms^{-1}$ and turbulence intensity $TI = 8.6\%$ at the turbine hub height as detailed in section~\ref{subsubsec:abl}.

This, combined with the spacing between the rows, means that the rows are independent of each other as is shown in previous studies~\citep{goriSensitivityAnalysisWake2023c, porteagelWindTurbineWindFarmFlows2020}.
When running the optimisation, we can therefore consider the angle of 10 turbines in a single row, as are highlighted in Fig.~\ref{fig:hornsrev}.
This significantly reduces the computational cost associated with running the LES\@.

\subsubsection{Single-Fidelity Optimisation}\label{subsubsec:10tsf}

\begin{figure}[htp]
    \begin{center}
        \includegraphics[width=\linewidth]{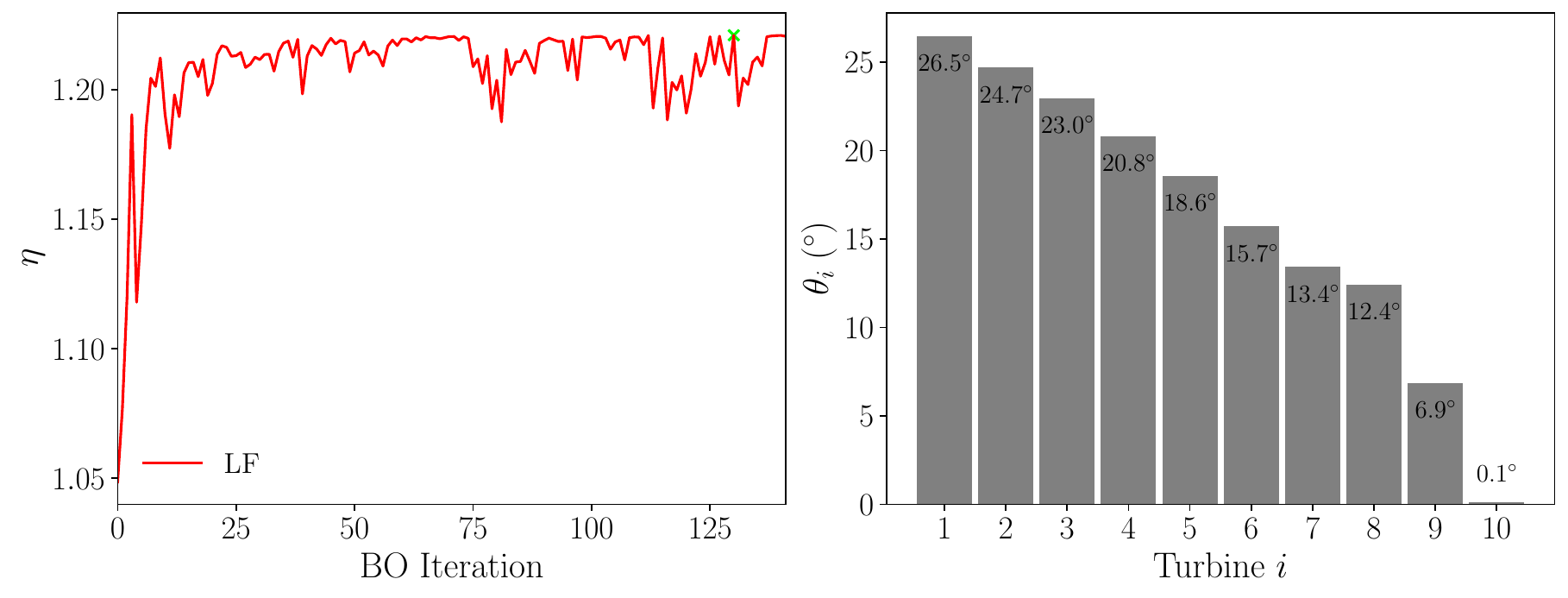}
        \caption{\textit{Left: }The progress of the wind farm efficency with BO iterations and \textit{Right: } the optimised yaw angles, using a GCH single-fidelity optimisation for the Horns Rev turbine array.}
        \label{fig:10tlf}
    \end{center}
\end{figure}

Firstly, a low-fidelity BO\@ is performed using the GCH wake model.
The initial dataset consists of a set of 100 yaw configurations selected using latin hypercube sampling and the corresponding calculated wind farm total power outputs.
For the optimisation, $\beta = 4$ and $\lambda = 1.0^{\circ}$ are used.
Fig.~\ref{fig:10tlf} shows the result of the optimisation with the power output obtained from the GCH wake model at each BO iteration showing the convergence criteria being met after 142 BO iterations, and the best result has a power improvement over the unoptimised configuration of $22.1\%$.
The optimised yaw angles are also shown in Fig.~\ref{fig:10tlf}, showing the first turbine yawed the most at $\theta_1 = 26.5^{\circ}$ and a mostly linear decrease in the yaw angle for downstream turbines until $\theta_1 = 0.1^{\circ}$.

\begin{figure}[htp]
    \begin{center}
        \includegraphics[width=\linewidth]{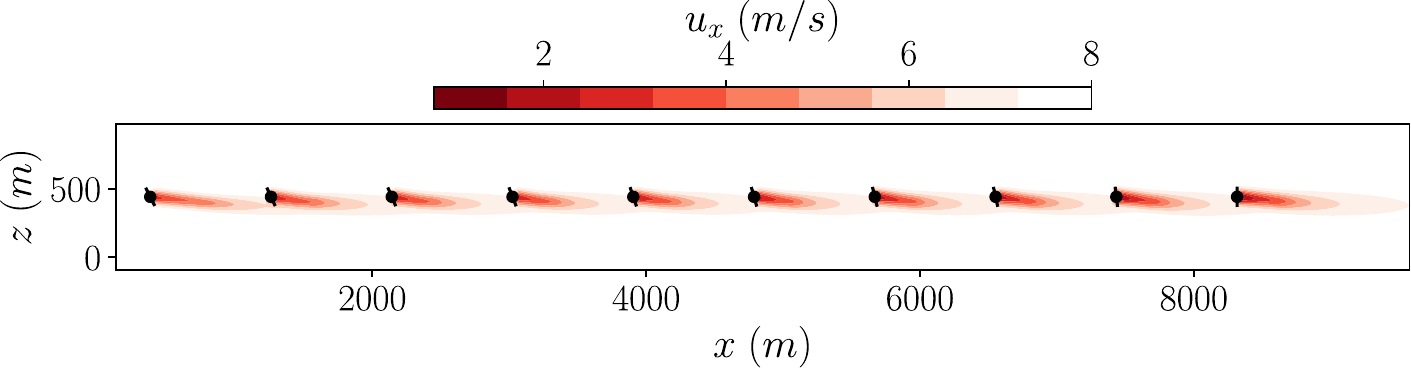}
        \caption{Optimised flow field at the turbine hub height using a GCH single-fidelity optimisation for the Horns Rev turbine array.}
        \label{fig:10tlf_slice}
    \end{center}
\end{figure}

Figure~\ref{fig:10tlf_slice} shows the flow field at the turbine hub height using the GCH wake model for the optimised yaw configuration.
It shows how, even as the yaw angles decrease downstream, the wakes are deflected away from subsequent turbines.
This is due to the secondary wake steering effects introduced in the GCH wake model (equation~\ref{eq:gch-wake}).

\begin{figure}[htp]
    \begin{center}
        \includegraphics[width=0.6\linewidth]{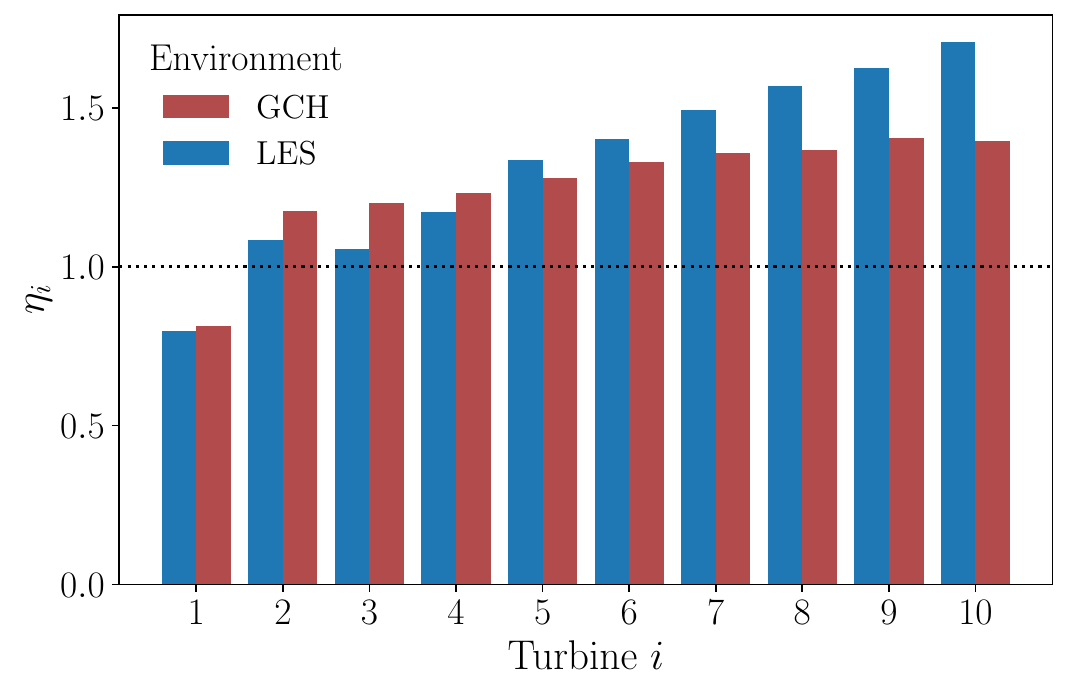}
        \caption{Turbine efficiency calculated in different environments for each turbine optimised using a GCH single-fidelity optimisation for the Horns Rev turbine array.}
        \label{fig:10tlf_powers}
    \end{center}
\end{figure}

The resulting wind farm yaw configuration was also tested in an LES environment for comparison.
In the LES environment the power improvement is $30.5\%$, showing that there is a discrepancy between the LES and wake model environments.
To investigate the cause of the discrepancy between environments, the individual turbine efficiencies calculated in the GCH and LES environments are shown in Figure~\ref{fig:10tlf_powers}.

The individual turbine efficiencies vary between the low- and high-fidelity environments.
The efficiency of the first turbine is similar between the two environments, with both showing less power generated at this turbine due to the yaw
This is offset by the increased power of the downstream turbines.
The wake steering appears to have a stronger effect in the wake model environment than the LES\@.
This leads to the turbines 2 to 4 having a higher efficiency in the wake model environment than in the LES\@.
Further downstream this effect diminishes, with turbines 5 to 10 having an increasingly higher efficiency in the LES than in the wake model environment.
This is likely due to the LES capturing the physics of the interacting turbulent wakes that the wake model is missing.

\subsubsection{Multi-Fidelity Optimisation}

\begin{figure}[htp]
    \begin{center}
        \includegraphics[width=\linewidth]{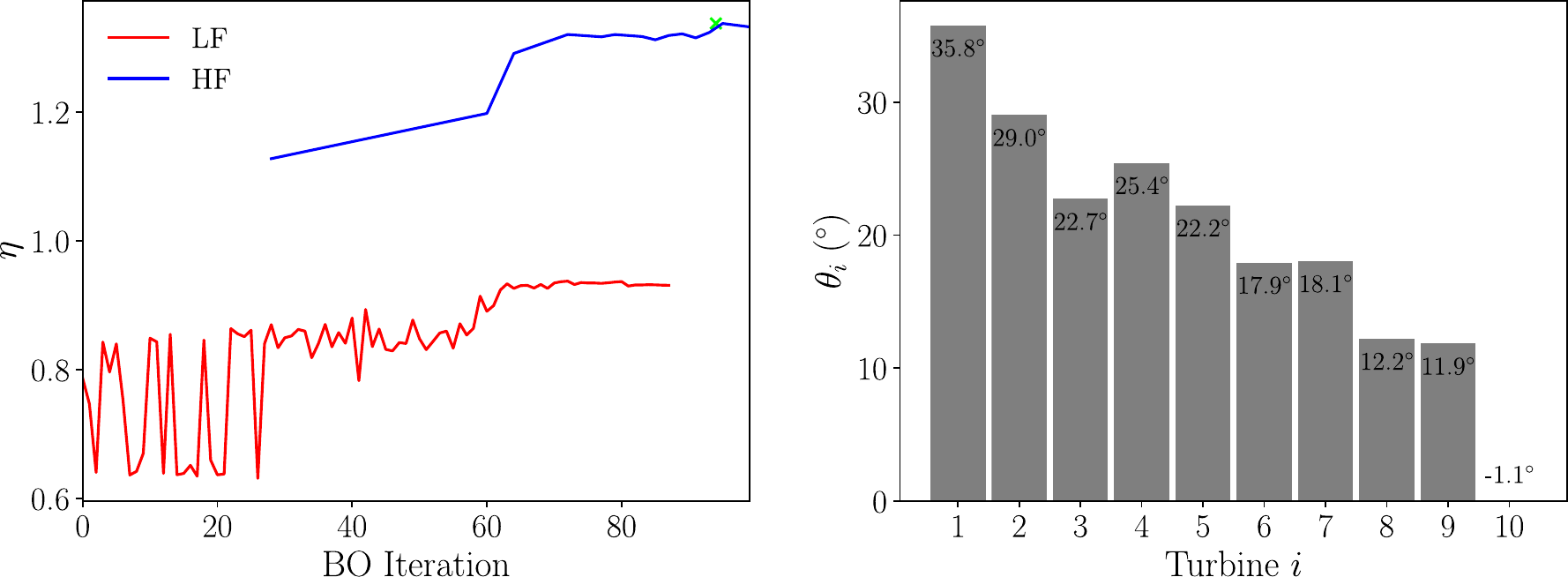}
        \caption{\textit{Left: }The progress of the wind farm efficency with BO iterations and \textit{Right: } The optimised turbine angles each turbine using a multi-fidelity optimisation for the Horns Rev turbine array.}
        \label{fig:10tmf_opt}
    \end{center}
\end{figure}

A multi-fidelity optimisation is performed on the same wind farm set-up with low-fidelity data provided by the GCH wake model and high-fidelity data from LES\@.
The initial datasets are generated using latin hypercube sampling with 100 low-fidelity and 12 high-fidelity yaw configurations and optimisation coefficients of $\beta_{HF} = 0.5$ and $\gamma_t^{LF} = 8$ are used with $\lambda = 2^{\circ}$.
This combination of optimisation coefficients gives a comparable low-fidelity exploration to the the single-fidelity optimisation in section~\ref{subsubsec:10tsf} whilst limiting the LES calculations.
The LES is performed on a uniform grid with dimensions $L_x \times L_y \times L_z = 74D \times 500m \times 7D$ and a grid spacing of $D/10$ in the three spatial directions.
The LES domain corresponds to the box shown in Fig~\ref{fig:hornsrev}, with periodic boundary conditions applied in the span-wise direction.
After an initialisation period, the power output from the LES calculation is averaged over a simulated time period of $\approx 2$ hours with a timestep of $0.2s$.
The results of the multi-fidelity optimisation are shown in Fig~\ref{fig:10tmf_opt} with the progress of the wind farm efficiency over the BO iterations for the two fidelities and the optimised yaw angle of each turbine.

The stopping criterion for the optimisation was met at 100 iterations with 86 low-fidelity iterations and 14 high-fidelity iterations.
This gives a power improvement of $33.8\%$, $3.3\%$ higher that using GCH wake model alone to perform the optimisation.
The optimised yaw angles differ from the single-fidelity optimisation, with the front turbines at a higher yaw angle with $\theta_1 = 35.8^{\circ}$ and $\theta_2 = 29^{\circ}$, and the turbines further downstream yawed to a comparable degree to in section~\ref{subsubsec:10tsf}.

\begin{figure}[htp]
    \begin{center}
        \includegraphics[width=\linewidth]{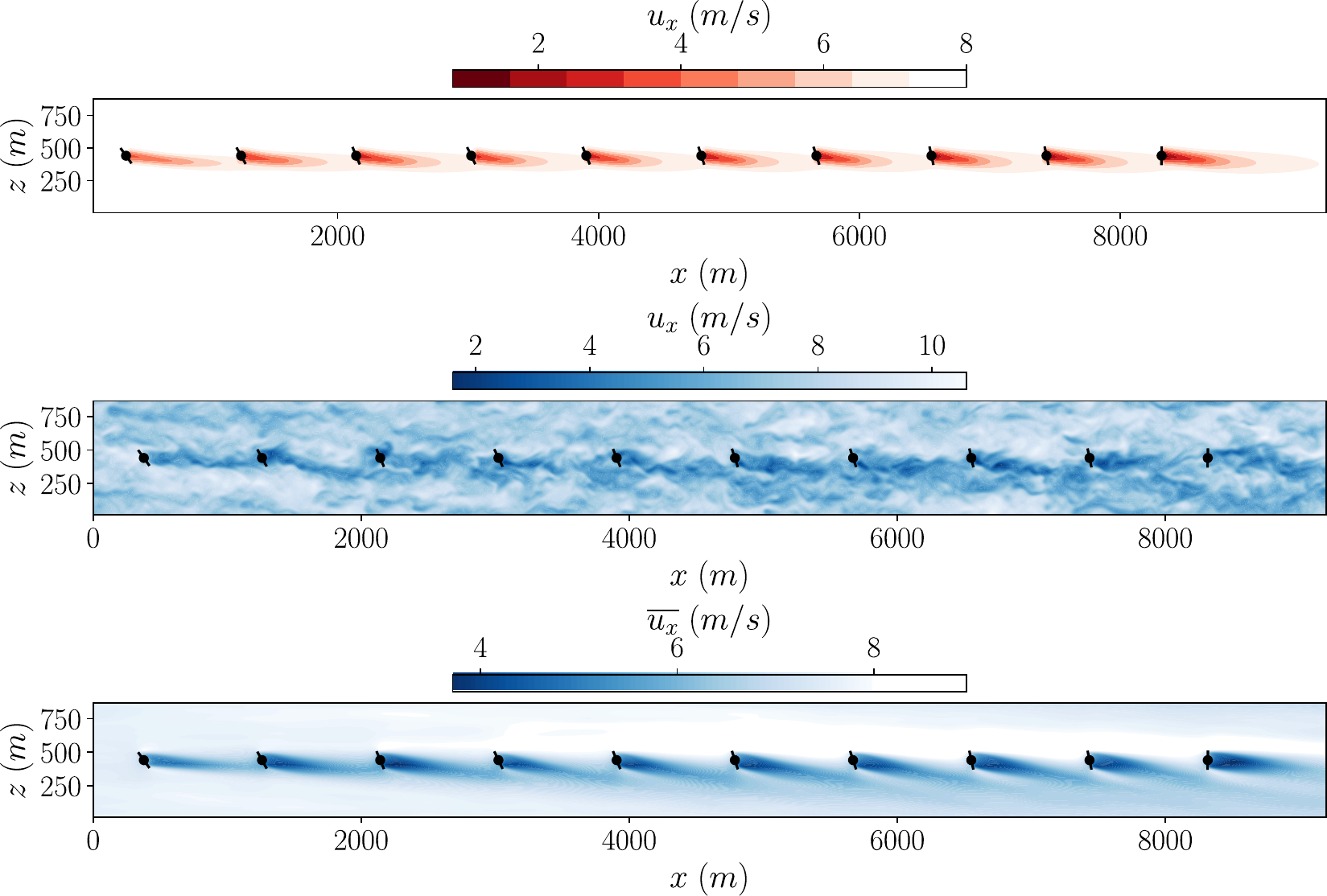}
        \caption{Slices of the \textit{Top: }GCH wake model, \textit{middle: }instantaneous LES and \textit{bottom: }mean LES stream-wise velocity fields for the multi-fidelity optimised yaw angles of the Horns-Rev wind farm.}
        \label{fig:10tmf_slice}
    \end{center}
\end{figure}

The optimised yaw configuration was simulated in both the LES and GCH wake model environments.
The resulting flow field at the turbine hub height is shown in Fig~\ref{fig:10tmf_slice} for both environments.

\begin{figure}[htp]
    \begin{center}
        \includegraphics[width=0.6\linewidth]{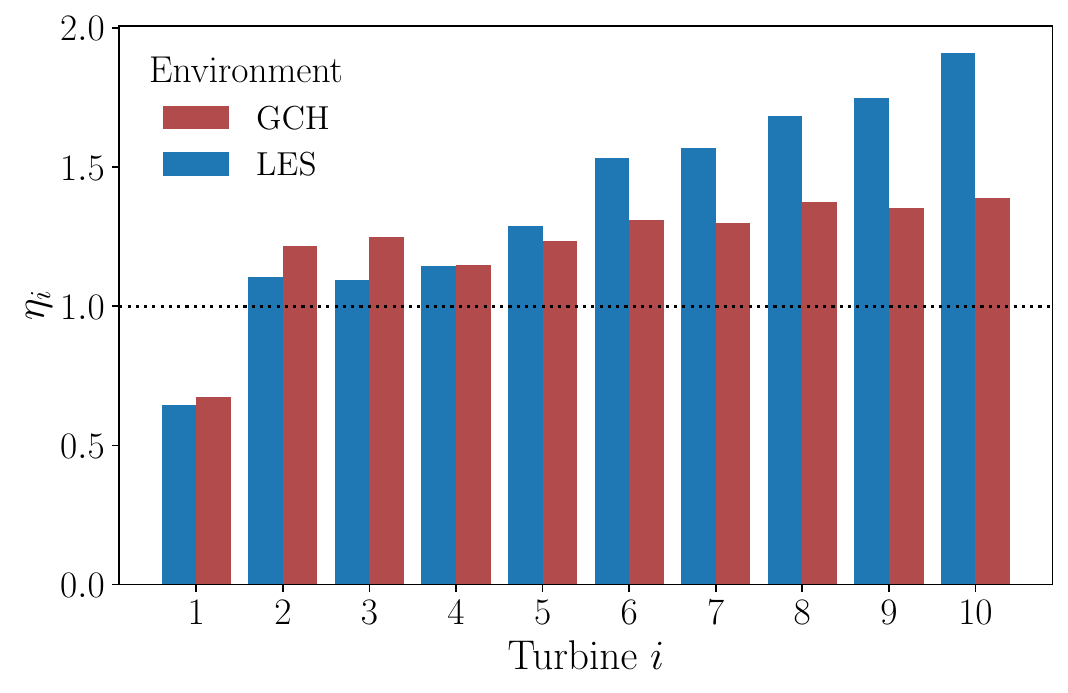}
        \caption{Turbine efficiency calculated in different environments for each turbine optimised using a multi-fidelity optimisation for the Horns Rev turbine array.}
        \label{fig:10tmf_powers}
    \end{center}
\end{figure}

Comparing the time averaged LES flow field to the GCH wake model flow field, the wakes behind the front three turbines are similar in both environments.
Further downstream, there is more difference in the wakes between the two environments with the LES wakes more deflected.
This results in a difference in power output for the downstream turbines between the environments.
This is shown in Fig.~\ref{fig:10tmf_powers} where the power efficiency of each wind turbine is shown for both environments.

Compared to the power output of the single-fidelity optimised case, the power of the first turbine is lower, as a result of the larger yaw angle.
However this is more than offset by the increased power output of the wind turbines 6 to 10.
This increased power output of these wind turbines is not captured by the GCH wake model, explaining why this configuration was not found by the single-fidelity optimisation.

\section{Conclusions} \label{sec:conclusions}

% What was done
In this study, we proposed a new multi-fidelity approach to optimising wind turbine yaw angles to maximise the power output from wind farms.
The multi-fidelity BO approach is based on a multi-fidelity NARGP model and a new MFUCB acquisition function.
For optimising the yaw configuration of wind farms, data was obtained during the optimisation from multiple fidelities of simulation using analytical wake models and LES\@.
This methodology was tested in a two-turbine wind farm and a larger wind farm representative of the Horns-rev wind farm.

% Results
When testing on the two-turbine wind farm, the multi-fidelity optimisation found an optimum yaw angle that was equivalent to the best high-fidelity optimisation result whilst reducing the computational cost by $65\%$ required to explore the parameter space.
When testing on a full wind farm, the LES and wake model environments provided different turbine power output results, suggesting that there are physical effects missing from the wake model that are captured by the LES\@.
The multi-fidelity optimisation using the GCH wake model as the low-fidelity and LES as the high-fidelity environments was able to exploit the additional physics captured to find an optimisation result with a higher wind farm power improvement of $33.8\%$.
The lower cost of running analytical wake models combined with the increase in accuracy of the LES simulations is exploited by the multi-fidelity optimisation to achieve improved optimisation results whilst limiting the computational expense.

% Future work
The approach could be extended in future studies by exploring different multi-fidelity surrogate models, such as multi-fidelity deep GPs within the optimisation.
Alternative multi-fidelity acquisition functions could be explored based on single-fidelity acquisition functions such as the expected improvement or probability of improvement acquisition functions.
Additionally a total budget for the optimisation could be included to vary the exploration/exploitation parameter as the optimisation progresses.
Including additional fidelities in the optimisation, such as RANS or actuator line model LES simulations could be explored within the framework presented here.

\subsection{Acknowledgements}
The authors gratefully acknowledge the support of the UK Engineering and Physical Sciences Research Council (EPSRC) for funding this research (EP/Y005619/1), and the UK Turbulence Consortium (EP/R029326/1 and EP/X035484/1).
The simulations were conducted on ARCHER2 and the computing time was awarded through an ARCHER2 Pioneer Project.

\subsection{Author Contributions}
A.M conducted the wing optimisation studies and simulation, A.M
prepared figures and wrote the main manuscript text.
S.L.\ contributed to the main manuscript text
and analysis of the results.
S.L.\ secured the funding.
All authors reviewed the manuscript.

\section{Declarations}
\subsection{Conflict of interest}
The authors have no conflict of interest.

\bibliography{Paper.bib}

\end{document}